\documentclass[pra,twocolumn,aps,showpacs]{revtex4-1}

\usepackage{amsmath}
\usepackage{amssymb}
\usepackage{hyperref}
\usepackage{graphicx}
\usepackage{color}
\usepackage{bm}
\everymath{\displaystyle}
\begin{document}

\title{Fluctuation and interaction induced instability of dark solitons in 
       single and binary condensates}

\author{Arko Roy}
\author{D. Angom}
\affiliation{Physical Research Laboratory,
             Navrangpura, Ahmedabad-380009, Gujarat,
             India}

\begin{abstract}
   We show that the presence of soliton in a single-species condensate, at
zero temperature, enhances the quantum depletion sufficient enough to induce 
dynamical instability of the system. We also predict that for two-species 
condensates, two Goldstone modes emerge in the excitation spectrum
at phase separation. Of these, one is due to the presence of the soliton.
We use Hartree-Fock-Bogoliubov theory with Popov approximation to examine 
the mode evolution, and demonstrate that when the anomalous mode
collides with a higher energy mode it renders the solitonic state oscillatory
unstable. We also report soliton induced change in the topology of the
density profiles of the two-species condensates at phase-separation.
\end{abstract}



\maketitle

\section{Introduction}
The experimental realization of single and multi-component Bose-Einstein
condensates (BECs) in atomic gases have opened up the possibility of exploring
topological defects. Due to the ubiquitous presence of topological defects
in nature, study of matter-wave excitations such as vortices and solitons in
atomic BECs has been a topic of extensive research both experimentally and
theoretically over the last few years. In fact, these have attracted much
attention as they are created spontaneously during BEC phase transition
through Kibble-Zurek mechanism ~\cite{,zurek_85,zurek_09,damski_10,
lamporesi_13}. A soliton, for instance can be used to probe the phase of 
the image acquired in a BEC interferometer as proposed by Negretti 
{\em et.  al}~\cite{negretti_04, negretti_08}. These and other
novel phenomena have inspired numerous experiments 
~\cite{burger_99,denschlag_00} and theoretical studies
~\cite{busch_00,busch_01,middelkamp-11,hoefer_11,kasamatsu_06,rajendran_09,
achilleos_12} with dark and bright solitons in atomic BECs in a wide range 
of settings under different scenarios. The experimental observation shows 
that the notch of the dark soliton gets filled up with thermal atoms over 
time and the soliton becomes gray, hence starts oscillating which are 
either short- or long-lived depending upon the system of
interest~\cite{becker_08,weller_08,stellmer_08}.

On the theoretical front, most of the studies on the statics and the dynamics
of dark solitons have been carried out in quasi-1D setting at zero temperature
where thermal fluctuations can be ignored ~\cite{muryshev_99}. There 
have been several works on stability of solitons in cigar-shaped double
well potential~\cite{middelkamp_10a}, disordered potential~\cite{mochol_12},
and optical lattice~\cite{kevrekidis_03,parker_04,sakaguchi_05,kartashov_11}. 
Stability of multiple solitons in quasi-1D trap has also been examined 
~\cite{atre_06,theocharis_10}.
Quantum depletion in BECs with soliton at $T=0$ in weakly interacting 
Bose gases has also been studied using approximate
models~\cite{dziarmaga-02,dziarmaga_02,dziarmaga_03,dziarmaga-03,law_02,
law_03,gangardt_10}. 
This motivated us to reexamine the role of
quantum fluctuations in BECs, whether it be
with or without soliton. We show that quantum fluctuation in BECs with
soliton is higher than without it. This is due to the presence of
the anomalous mode, and we demonstrate that quantum fluctuations can
make the dark soliton gray, which as a result becomes dynamically unstable.

The two-component BECs (TBECs), on the other have different ground states
depending on the interactions, as compared to a single-component BEC. The
most unique aspect of TBECs is the phenomenon of phase separation. 
Most importantly, in experiments, the TBEC
can be steered from miscible to phase-separated domain or vice versa through a
Feshbach resonance~\cite{papp_08,tojo_10}. This has motivated numerous 
theoretical investigations on
stationary states~\cite{ho_96,gautam-10,gautam-11}, dynamical
instabilities~\cite{sasaki_09,gautam_10,kadokura-12}, 
and collective
excitations~\cite{takeuchi_13,ticknor_13,roy_14,ticknor_14,mason_14} 
of TBEC. Furthermore, repulsive TBECs support coupled dark-bright solitons 
which makes it richer and more interesting than single-component 
BECs~\cite{busch_01}. The bright soliton, on the other hand, 
cannot survive in single component BECs with repulsive interaction. It may
be mentioned here that, solitons in BECs and TBECs have been experimentally 
achieved either by phase-imprinting method~\cite{becker_08} or 
in two counter-flowing miscible TBECs above a critical 
velocity~\cite{hamner_11}. For miscible TBECs, creation and interaction of 
dark solitons has been theoretically examined in 
Refs.~\cite{ohberg_01,ohberg_01a}. Families of stable solitonic 
solutions from coupled Gross-Pitaevskii(GP) equations in quasi-1D TBECs at zero
temperature have been obtained ~\cite{kevrekidis_04,schumayer_04}. 

In the present work, we describe the development
of Hartree-Fock-Bogoliubov theory with Popov (HFB-Popov) approximation for
trapped TBECs. We use it to examine the evolution of Goldstone modes and
mode energies for TBEC with soliton as a function of interspecies
scattering length. Recent works ~\cite{ticknor_13,takeuchi_13,roy_14,ticknor_14}
have reported the existence of an additional Goldstone mode at phase 
separation in the symmetry-broken density profiles. We have demonstrated in 
our earlier work~\cite{roy_14} that in the {\em sandwich type} density profiles 
where one of the species is surrounded on both sides by the other, 
the mode evolves very differently with the
appearance of a third Goldstone mode. In the present work, we
show the presence of the soliton introduces an additional Goldstone mode to
the system. Even at zero temperature without considering any quantum 
fluctuation, for certain range of
interspecies scattering length, the TBEC becomes dynamically unstable. The
difference in the mass of the two species also plays a significant role in mode
evolution and topology of density profiles.



\section{Theory}
\subsection{Single component BEC}
\label{theorya}
 For a quasi-1D system, the
trapping frequencies in 
$V=(1/2)m(\omega_x^2x^2 + \omega_y^2y^2 + \omega_z^2z^2)$ should satisfy 
the condition 
$\omega_x=\omega_y=\omega_{\perp} \gg \omega_z $. The condensate wave
function 
in such a potential can be integrated out along $xy$ direction to reduce it 
to a quasi-1D system.
The grand-canonical Hamiltonian, in second quantized form, describing an
interacting BEC is then
\begin{eqnarray}
  H &=& \int dz\hat{\Psi}^\dagger(z,t)
        \bigg[-\frac{\hbar^{2}}{2m}\frac{\partial ^2}{\partial z^2} 
        + V(z)-\mu\nonumber\\ 
    & & + \frac{U}{2}\hat{\Psi}^\dagger(z,t)\hat{\Psi}
        (z,t)\bigg]\hat{\Psi}(z,t),
\label{hamiltonian} 
\end{eqnarray}
where  $\hat{\Psi}$ is the Bose field 
operator of the single species BEC, and $\mu$  is the chemical 
potential.
The strength of the intra-species repulsive interactions is
$U = (a\lambda)/m$ ,  where
$\lambda = (\omega_{\perp}/\omega_z) \gg 1$ is the anisotropy parameter,
$a$ is the $s$-wave scattering length, $m$ is the atomic mass of 
the species. Starting with this Hamiltonian, the equation of motion of the 
Bose field operator is 
\begin{eqnarray}
i\hbar\frac{\partial}{\partial t}\hat{\Psi}=
\hat{h}\hat{\Psi} + U\hat{\Psi}^\dagger\hat{\Psi}\hat{\Psi},
\end{eqnarray}
where $\hat{h}= (-\hbar^{2}/2m)\partial ^2/\partial z^2
+V(z)-\mu$. 
For the sake of simplicity of notation, we will refrain from writing the 
explicit dependence of $\hat{\Psi}$ on $z$ and $t$. Since a majority of 
the atoms populate the ground state for the temperature domain pertinent to the 
experiments ($T\leqslant 0.65T_c$ ) \cite{dodd_98}, the condensate part
can be separated out from the Bose field operator
$\hat{\Psi}(z,t)$. 
The non-condensed or the thermal cloud of atoms are then the fluctuations
of the condensate field. Here, $T_c$ is the critical temperature of ideal
gas in a harmonic confining potential. Accordingly, we define 
\cite{griffin_96}, $\hat{\Psi}(z,t) = \phi(z,t) + \tilde\psi(z,t)$, 
where $\phi(z, t)$  is a $c$-field and represents the condensate, and 
$\tilde\psi(z,t) $ is the fluctuation part. For a single component BEC, 
$\hat{\Psi}$ can then be written as 
\begin{eqnarray}
\hat{\Psi} = \phi + \tilde\psi.
\end{eqnarray}
Thus for a single-species BEC, the equation of motion of the 
condensate within the time-independent HFB-Popov approximation is given by 
the generalized GP equation
\begin{equation}
  \hat{h}\phi + U\left[n_{c}+2\tilde{n}\right]\phi = 0.
  \label{gpe1s}
\end{equation}    
In the above equation,
$n_{c}(z)\equiv|\phi(z)|^2$,
$\tilde{n}(z)\equiv\langle\tilde{\psi}^{\dagger}(z,t)
\tilde{\psi}(z,t)\rangle$, and $n(z) = n_{c}(z)+ \tilde{n}(z)$
are the local condensate, non-condensate, and total density,
respectively. Using Bogoliubov transformation, the fluctuations are
\begin{eqnarray}
   \tilde{\psi}(z,t) &=&\sum_{j}\left[u_{j}(z)
    \hat{\alpha}_j(z) e^{-iE_{j}t}
   - v_{j}^{*}(z)\hat{\alpha}_j^\dagger(z) e^{iE_{j}t}
    \right], \nonumber \\
    \tilde{\psi}^\dagger(z,t)
    &=&\sum_{j}\left[u_{j}^*(z)\hat{\alpha}_j^\dagger(z)e^{iE_{j}t}-
    v_{j}(z)\hat{\alpha}_j(z) e^{-iE_{j}t}\right].\nonumber
\label{ansatz}
\end{eqnarray}
Here, $\hat{\alpha}_j$ ($\hat{\alpha}_j^\dagger$) are the quasiparticle
annihilation (creation) operators and satisfy the usual Bose commutation 
relations, and the subscript $j$ represents the energy eigenvalue index.
From the above definitions, we get the following Bogoliubov-de Gennes 
( BdG) equations
\begin{subequations}
\begin{eqnarray}
(\hat{h}+2Un)u_{j}-U\phi^{2}v_{j}&=&E_{j}u_{j},\\
-(\hat{h}+2Un)v_{j}+U\phi^{*2}u_{j}&=& E_{j}v_{j}.
\end{eqnarray}
\label{bdg1}
\end{subequations}
The number density $\tilde{n}$ of non-condensate particles is then
\begin{equation}
 \tilde{n}=\sum_{j}\{[|u_{j}|^2+|v_{j}|^2]N_{0}(E_j)+|v_{j}|^2\},
  \label{n_tilde}
\end{equation}
where $\langle\hat{\alpha}_{j}^\dagger\hat{\alpha}_{j}\rangle = (e^{\beta
E_{j}}-1)^{-1}\equiv N_{0}(E_j)$ with $\beta=1/k_{\rm B} T$, is the Bose 
factor of the quasi-particle state with energy $E_j$ at temperature $T$.  
However, it should be emphasized
that, when $T\rightarrow0$, $ N_0(E_j)$'s in Eq. (\ref{n_tilde}) vanishes.
The non-condensate density is then reduced to
\begin{equation}
 \tilde{n} = \sum_{j}|v_{j}|^2.
 \label{n_r}
\end{equation}
Thus, at zero temperature we need to solve the equations self-consistently
as the quantum depletion term $|v_{j}|^2$ in the above equation is non-zero.


\subsection{Harmonic oscillator basis}
\label{harmonic}
We solve the quasi-particle amplitudes $u_j$, $v_j$'s in the basis of the 
harmonic oscillator trapping potential.
\begin{eqnarray}
 u_{j} = \sum_{i=0}^{N_b} p_{ij}\xi_i,\;\;
 v_{j} = \sum_{i=0}^{N_b} q_{ij}\xi_i,\nonumber \\
\label{exp}
\end{eqnarray}
where $\xi_i$ is the $i{\rm th}$ harmonic oscillator eigenstate
and $N_b$ is the number of basis that is considered.
Using this expansion, Eq. (\ref{bdg1}) is then
reduced to a matrix eigenvalue equation and solved using standard matrix
diagonalization algorithms.  The matrix has a dimension of
$2N_b\times2N_b$, and
is non-Hermitian, non-symmetric and may have complex eigenvalues.
The eigenvalue spectrum obtained from
the diagonalization of the matrix has an equal number of positive and
negative eigenvalues $E_j$'s. In addition, the amount of energy that is
carried by the eigenmode $j$ is given by
\begin{equation}
 \Delta_j = \int\,dz(|u_j|^2 - |v_j|^2)E_j.
 \label{kr_sign}
\end{equation}
The sign of the quantity $\Delta_j$ is known as {\em Krein sign}. If this
sign turns out to be negative for a mode $j$, then the corresponding mode is
called as the {\em anomalous mode}. It signifies the energetic
instability which may be present due to a topological defect
in the system.


\subsection{Hartree-Fock basis}
\label{hfock}
To incorporate the interactions present in the system while calculating the
Bogoliubov quasi-particle amplitudes $u_j$ and $v_j$'s more efficiently, in
terms of basis size, we resort to {\rm Hartree-Fock basis}.
Thus, to solve Eq. (\ref{bdg1}), we define $u_j$'s and $v_j$'s as a linear
combination of {\em Hartree-Fock basis} functions $\zeta_k$,
\begin{eqnarray}
u_j = \sum_k c_k^j\zeta_k, v_j = \sum_k d_k^j\zeta_k,
\label{decomp}
\end{eqnarray}
 where $c_k$, and $d_k$ are the coefficients of linear combination. In
 principle, the GP equation has an infinite number of eigenvalues
 $\epsilon_k$ and eigenvectors $\zeta_k$. In general, Eq. (\ref{gpe1s}) can 
 then be recast into a matrix eigenvalue equation
 \begin{equation}
  {\mathcal H}\zeta_k = \epsilon_k\zeta_k,
  \label{egveq}
 \end{equation}
 where ${\mathcal H} = {\hat h} + U[n_c+2\tilde{n}]$, and $k$ stands for
 the eigenvalue index. The eigensolution with the lowest eigenvalue
 $\epsilon_0$ is referred to as the condensate ground state with the
 condensate wave function $\phi\equiv\zeta_0$. 
 To calculate the quasi-particle amplitudes $u_j$'s and $v_j$'s we again 
 expand the eigensolutions $\zeta_k$ in terms of $\xi_i$, then 
 \begin{equation}
  \zeta_k = \sum_i a_i^k\xi_i,
  \label{decomphf}
 \end{equation}
 Taking the orthogonality and linear independence of $\xi_i$s into account
 and plugging Eq. (\ref{decomphf}) in Eq. (\ref{egveq}), one can obtain the
 expansion coefficients $a_k$ used in decomposing the above
 equation. This yields a set of basis functions $\{\zeta_k\} $, which is 
generally referred to as the {\em Hartree-Fock basis}. The choice of 
$\zeta_k$ reduces the number of basis functions required in the calculation 
of $u_j$ and $v_j$'s as $\zeta_k$ subsumes the effect of interactions in the 
system.

\subsection{Two component BEC}
\label{theoryb}
Similarly, for a TBEC in a quasi-1D trapped system,
\begin{eqnarray}
  H &=& \sum_{k=1,2}\int dz\hat{\Psi}_{k}^\dagger(z,t)
          \bigg[-\frac{\hbar^{2}}{2m_k}\frac{\partial ^2}{\partial z^2} 
    + V_k(z)-\mu_k\nonumber\\ 
    & & +
    \frac{U_{kk}}{2}\hat{\Psi}_{k}^\dagger(z,t)\hat{\Psi}_{k}(z,t)\bigg]
    \hat{\Psi}_{k}(z,t)\nonumber\\ 
    & & + U_{12}\int dz
   \hat{\Psi}_{1}^\dagger(z,t)\hat{\Psi}_{2}^\dagger(z,t)
    \hat{\Psi}_{1}(z,t)\hat{\Psi}_{2}(z,t),
\label{hamiltonian} 
\end{eqnarray}
where $k=1,2$ is the species index, $\hat{\Psi}_k$'s are the Bose field 
operators of the two different species, and $\mu_k$'s  are the chemical 
potentials.
The strength of intra and inter-species repulsive interactions are 
$U_{kk} = (a_{kk}\lambda)/m_{k}$ 
and $U_{12}=(a_{12}\lambda)/(2m_{12})$, respectively, where
$\lambda = (\omega_{\perp}/\omega_z) \gg 1$ is the anisotropy parameter,
$a_{kk}$ is the $s$-wave scattering length, $m_k$'s are the atomic masses
of 
the species and $m_{12}=m_1 m_2/(m_1+m_2)$. Starting with this
Hamiltonian, the equation of motion of the Bose field operators is 
\begin{equation}
 i\hbar\frac{\partial}{\partial t}
  \begin{pmatrix}
     \hat{\Psi}_1\\
        \hat{\Psi}_2
        \end{pmatrix} \!\!
        = \!\!\begin{pmatrix}
    \hat{h}_1 + U_{11}\hat{\Psi}_1^\dagger\hat{\Psi}_1 & U_{12}
    \hat{\Psi}_2^\dagger \hat{\Psi}_1\\
    U_{12}\hat{\Psi}_1^\dagger\hat{\Psi}_2
    & \hat{h}_2
    + U_{22}\hat{\Psi}_2^\dagger\hat{\Psi}_2
    \end{pmatrix} \!\!\!
 \begin{pmatrix}
    \hat{\Psi}_1\\
    \hat{\Psi}_2  
 \end{pmatrix}, \nonumber
 \label{twocomp}
\end{equation}
where $\hat{h}_{k}= (-\hbar^{2}/2m_k)\partial^2/\partial z^2+V_k(z)-\mu_k$.
In the same way as in single species case, we define 
\cite{griffin_96}, $\hat{\Psi}(z,t) = \Phi(z) + \tilde\Psi(z,t)$, 
where $\Phi(z)$  is a $c$-field and represents the condensate, and 
$\tilde\Psi(z,t) $ is the fluctuation part.
In two component representation
\begin{eqnarray}
\begin{pmatrix}
 \hat{\Psi}_1\\
 \hat{\Psi}_2
\end{pmatrix}
   =
\begin{pmatrix}
 \phi_1\\
 \phi_2
\end{pmatrix}
   +
\begin{pmatrix}
 \tilde\psi _1\\
 \tilde\psi _2\\
\end{pmatrix},
\end{eqnarray}
where $\phi_k(z)$ and $\tilde{\psi}_k(z)$ are the condensate and fluctuation 
part of the $k$th species. Thus for a TBEC, $\phi_k$s are the 
stationary solutions of the coupled generalized GP equations, with 
time-independent HFB-Popov approximation, given by
\begin{subequations}
\begin{eqnarray}
  \hat{h}_1\phi_1 + U_{11}\left[n_{c1}+2\tilde{n}_{1}\right]\phi_1
    +U_{12}n_2\phi_1=0,\\
      \hat{h}_2\phi_2 + U_{22}\left[n_{c2}+2\tilde{n}_{2}\right]\phi_2
    +U_{12}n_1\phi_2=0.
\end{eqnarray}
\label{gpe}
\end{subequations}
In the above equation,
$n_{ck}(z)\equiv|\phi_k(z)|^2$,
$\tilde{n}_k(z)\equiv\langle\tilde{\psi}_{k}^{\dagger}(z,t)
\tilde{\psi}_k(z,t)\rangle$, and $n_k(z) = n_{ck}(z)+ \tilde{n}_k(z)$
are the local condensate, non-condensate, and total density,
respectively. Using Bogoliubov transformation, the fluctuations are
\begin{eqnarray}
   \tilde{\psi}_k(z,t) &=&\sum_{j}\left[u_{kj}(z)
    \hat{\alpha}_j(z) e^{-iE_{j}t}
    -v_{kj}^{*}(z)\hat{\alpha}_j^\dagger(z) e^{iE_{j}t}
    \right], \nonumber \\
\tilde{\psi}_{k}^\dagger(z,t)
&=&\sum_{j}\left[u_{kj}^*(z)\hat{\alpha}_j^\dagger(z)e^{iE_{j}t}-v_{kj}(z)
\hat{\alpha}_j(z) e^{-iE_{j}t}\right].
\nonumber
\label{ansatz}
\end{eqnarray}
From this formalism we obtain the following BdG equations
\begin{subequations}
\begin{eqnarray}
 \hat{{\mathcal L}}_{1}u_{1j}-U_{11}\phi_{1}^{2}v_{1j}+U_{12}\phi_1 \left 
   (\phi_2^{*}u_{2j} -\phi_2v_{2j}\right )&=& E_{j}u_{1j},\;\;\;\;\;\;\\
    \hat{\underline{\mathcal L}}_{1}v_{1j}+U_{11}\phi_{1}^{*2}u_{1j}-U_{12}
    \phi_1^*\left (\phi_2v_{2j}-\phi_2^*u_{2j} \right ) 
     &=& E_{j}v_{1j},\;\;\;\;\;\;\\
    \hat{{\mathcal L}}_{2}u_{2j}-U_{22}\phi_{2}^{2}v_{2j}+U_{12}\phi_2\left 
    ( \phi_1^*u_{1j}-\phi_1v_{1j} \right ) &=& E_{j}u_{2j},\;\;\;\;\;\;\\
\hat{\underline{\mathcal L}}_{2}v_{2j}+U_{22}\phi_{2}^{*2}u_{2j}-U_{12} 
\phi_2^*\left ( \phi_1v_{1j}-\phi_1^*u_{1j}\right ) &=& 
E_{j}v_{2j},\;\;\;\;\;\;\;\;\;
\end{eqnarray}
\label{bdg2}
\end{subequations}
where $\hat{{\mathcal L}}_{1}=
\big(\hat{h}_1+2U_{11}n_{1}+U_{12}n_{2})$, $\hat{{\mathcal L}}_{2}
=\big(\hat{h}_2+2U_{22}n_{2}+U_{12}n_{1}\big)$ and 
$\hat{\underline{\cal L}}_k  = -\hat{\cal L}_k$. The number density
$\tilde{n}_k$ of non-condensate particles is then
\begin{equation}
 \tilde{n}_k=\sum_{j}\{[|u_{kj}|^2+|v_{kj}|^2]N_{0}(E_j)+|v_{kj}|^2\},
  \label{n_k2}
  \end{equation}
To solve Eq. (\ref{bdg2}) we define $u$ and $v$'s as linear
combination of $\xi_i$s. The equation is then
reduced to a matrix eigenvalue equation and solved using standard matrix
diagonalization algorithms.


\section{The Dark soliton}
\label{soliton}
The location of a dark soliton is a place in a quasi-1D condensate is where
the condensate wave function $\phi(z)$ changes sign. The condensate wave 
function then has a kink where the density is zero. Typically, a wave 
function of the dark
soliton is simply proportional to $\tanh[(z-z_0)/\xi]$, where $\xi$ is a
local value of the healing length at position $z_0$ of the soliton.
Hereafter, it is to be noted that the symbol $\xi$ without any subscript
refers to the healing length.
Condensate with a soliton at $z_0=0$ is an antisymmetric wave function of
$z$ and the phase of the wave function jumps discontinuously by $\pi$ as
$z$ passes through zero. Even at $T=0$, quantum depletion from the
condensate leads to graying of the dark soliton. The kink of the soliton
gets filled up with incoherent atoms quantum depleted from the condensate.
The soliton is created by employing phase-imprinting method~\cite{becker_08}. 
We assume that
before phase imprinting all the atoms of the system is in symmetric ground
state. Right after this operation one gets a condensate with an
antisymmetric wave function.


\section{Results and Discussions}
\label{numerical}
\subsection{Numerical Details}

For single component BECs at  $T=0$ studies we solve
Eq.~\ref{gpe1s} neglecting the non-condensate density ($\tilde{n}=0$)
using finite-difference methods and in particular, we use the split-step
Crank-Nicholson method ~\cite{muruganandam_09}. For TBECs, we proceed in a 
similar way by solving the pair of coupled Eqs.~(\ref{gpe}) and 
setting $\tilde{n}_k=0$. The method when implemented with imaginary time 
propagation is appropriate to obtain the stationary ground state wave 
function of the single component BEC or TBEC. Furthermore, we use numerical
implementation of the phase imprinting method to generate a dark soliton in
(T)BEC. For this, we begin the simulation with imaginary time propagation
of the GP equation and imprint $\pi$ phase jump corresponding to a soliton 
at $z_0=0$ by 
using $\phi = |\phi|\exp(i\pi)$. Using this solution of the GPE, and 
based on Eq.~(\ref{exp}), we cast the Eq. (\ref{bdg1})as a matrix eigenvalue 
equation in the basis of the trapping potential. 
The matrix is then diagonalized using the LAPACK routine
{\tt zgeev} \cite{anderson_99} to find the quasi-particle energies and 
amplitudes, $E_j$, and $u_j$'s and $v_j$'s, respectively. 
We begin our $T=0$ calculations to account for quantum fluctuations with
this step. This sets the starting point of the first iteration where 
the $u_j$'s and $v_j$'s along with positive energy modes $E_j$'s are used
to get the initial estimate of $\tilde{n}$ through Eq.~(\ref{n_tilde}). The
ground state wave function $\phi$ of  BEC and chemical potential $\mu$ are 
again re-determined from Eq.~(\ref{gpe1s}), using this updated value of
$\tilde{n}$. For calculation of eigenmodes of TBEC with soliton, we again 
cast Eq. (\ref{bdg2}) as a matrix and diagonalize it~\cite{roy_14}. During the 
calculation of the $u_k$ and $v_k$, we choose an 
optimal number of the harmonic oscillator basis functions. 


\subsection{Single species BEC}
The low-lying excitation spectrum of a quasi-1D BEC with a soliton
is characterized by the presence of an anomalous mode, which indicates that
the BEC is in an energetically excited state. This is in
addition to the Goldstone and the Kohn modes, which are also present in the
excitation spectrum of a quasi-1D BEC without soliton. 
The anomalous, and Kohn mode energies are real, and the energy of the 
anomalous mode $\approx \hbar\omega_z/\sqrt{2}$. A unique feature of 
the anomalous mode is the negative 
\emph {Krein sign} \cite{dziarmaga-02,middelkamp_10a}. This shows 
the solitonic solution of the stationary quasi-1D GP equation is stable. 
However,  when the solution is evolved in imaginary time, with the inclusion 
of $\tilde{n}$ in the $T=0$ GP equation, the anomalous mode is 
transformed into an imaginary energy eigenmode. 
This is an unambiguous signature of 
quantum depletion induced instability of the solitonic solution. 
In other words, the non-zero $\tilde{n}$ arising from the quantum 
fluctuations within the notch of the soliton turns it 
gray, and renders the system dynamically unstable. Furthermore, the low-lying 
energy spectrum is devoid of any negative \emph {Krein sign} 
eigenmodes. The anomalous mode, however, reappears in the excitation 
spectrum on further evolving the system over imaginary time.
\begin{figure}[h]
 \includegraphics[width=8.5cm]{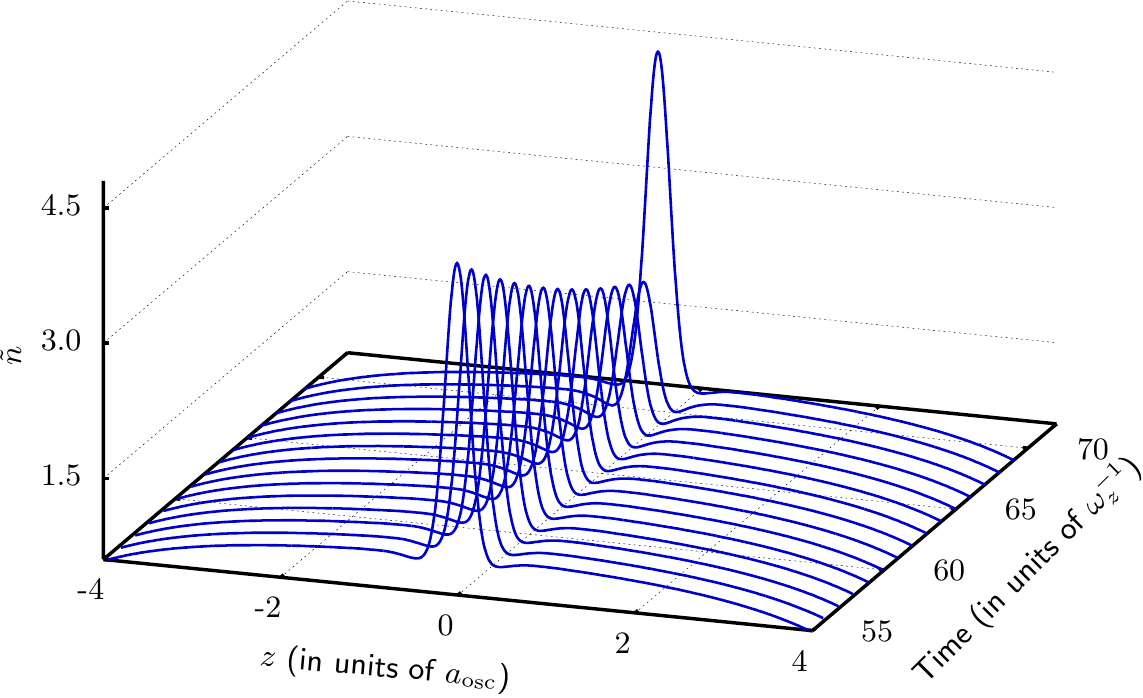}
 \caption{The temporal evolution in the profile of the non-condensate atom 
          density $\tilde n$ at $T=0$ measured in units of $a_{\rm osc}^{-1}$, 
          where $a_{\rm osc} = \sqrt{\hbar/(m\omega_z)}$. The plots show a
          steady drop in the number of non-condensate atoms till it reaches
          a threshold value, and then, the anomalous mode reappears in the 
          spectrum. The latter is reflected in the profile of $\tilde n$ at 
          $t = 69 \omega_z^{-1}$, where it has maximal distribution.
         }
\label{density_iteration}         
\end{figure}

To further examine the trend in the evolution of $E_{\rm an}$, the energy of 
the anomalous mode or the first excited state, we study the variation of 
$\tilde{n}$ with time as  shown in Fig.~\ref{density_iteration}. 
The contribution from the anomalous mode fills up the notch of the 
soliton and $\tilde{n}(0)$ has the largest possible value at the initial
state of evolution. At later times, 
$E_{\rm an}$ is imaginary and $\tilde{n}(0)$ decreases, the trend is as 
shown in Fig.~\ref{density_iteration}. However, when $\tilde{n}(0)$ reaches 
a critical value, which in the present work 
is $\approx 2.312 ~ a_{\rm osc}^{-1}$, it is no longer large enough to render 
the solitonic solution unstable and the anomalous mode reappears. This confirms 
$\tilde{n}(0)$ has a threshold value below which the solitonic solution may
be stable.

\begin{figure}[h]
 \includegraphics[height=5.0cm]{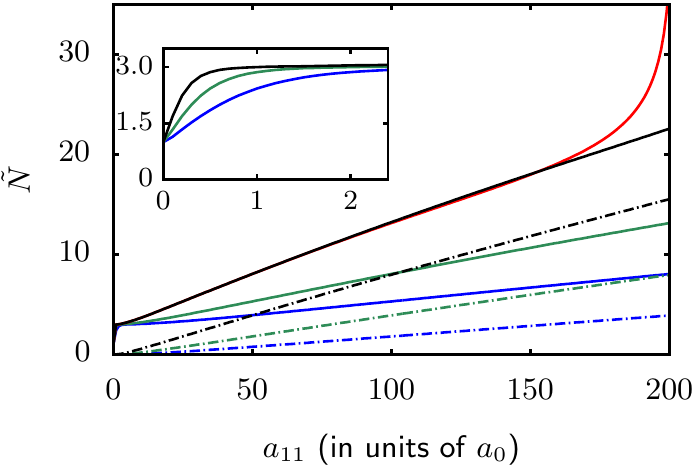}
  \caption{Variation in the total number of non-condensate atoms $\tilde N$ 
           at $T=0$ as a function of the scattering length $a_{11}$. The 
           solid (dashed) blue, green, and black lines represent $\tilde N$ 
           in the presence (absence) of soliton with total number of atoms
           $N$ = 500, 1000, and 2000, respectively. The solid red line
           represents $\tilde N$ in the presence of soliton for $N = 2000$, 
           with the number of basis $N_b=170$, it is shown to indicate lack of
           accuracy at higher $a_{11}$ with lower number of basis functions. 
           The inset plots show the trend of $\tilde{N}$ in the neighbourhood 
           of $a_{11}\approx 0$, where there is a sharp increase.
           }
\label{quant_depletion}               
\end{figure}

For the limiting case of $a_{\rm RbRb} \rightarrow 0$, or the non-interacting 
limit the Bogoliubov modes are, to a very good approximation, the 
eigenstates of the trapping potential. In this 
limit too, the condensate with the soliton has higher $\tilde{n}$ than
the condensate without soliton. An exponential increase in the total 
number of non-condensate atoms
\begin{equation}
  \tilde{N} = \int_{-\infty}^{\infty} \tilde{n}~dz,
\end{equation}
is observed
as $a_{\rm RbRb}$ is increased from near-zero to $a_{\rm RbRb}\approx a_0$, 
this is evident from the inset plot in Fig.~\ref{quant_depletion}. However, 
$\tilde{N}$ increases linearly with further increase of $a_{\rm RbRb}$ and this 
is shown in the main plot of Fig.~\ref{quant_depletion}. 
An important observation is that, $d\tilde{N}/da_{\rm RbRb}\propto N$  
(total number of atoms), which is due to higher repulsive 
interaction energy with increasing $N$. This is visible in the family of 
curves given for different values of $N$ in Fig.~\ref{quant_depletion}.
It should be emphasized here that an optimal choice of basis size $N_b$ is
necessary in all the computations to obtain accurate mode functions and
energies. For weakly interacting condensates with soliton,  a 
basis set consisting of 170 basis functions give converged and reliable 
results. But, for the strongly interacting case $1\ll NU$, the energy 
eigenvalues $E_j$s do not converge and $\tilde{N}$ diverges as shown by the 
red solid line in Fig.~\ref{quant_depletion} for $N=2000$. However, we get 
converged and reliable results when the basis size is increased to 240 
basis functions. 

The results that we have presented in this  section correspond to a condensate 
with a soliton at the center of the trap consisting of $N = 2000$ $^{87}$Rb 
atoms whose $s$-wave scattering length is $a_{11} = a_{\rm RbRb} = 100a_0$, 
where $a_0$ is the Bohr radius. The 
evolution of the low-lying modes are computed for the above-mentioned 
$a_{\rm RbRb}$ with $\omega_z$ = $2\pi\times 4.55$Hz, 
and $\omega_\perp = 20\omega_z$. This
choice of parameters are consistent with the experimental setting and 
satisfies the condition of quasi-1D approximation 
\cite{gorlitz_01, weller_08, pattinson_13}.
  It must be mentioned here that, we get almost identical results using 
either the harmonic oscillator basis or the {\em Hartree-Fock} basis. With the 
latter, in general, we require a smaller basis size. However, for the present 
work on quasi-1D condensates, the dimension of the BdG matrix  is within 
manageable limits even with the harmonic oscillator basis. 

\begin{figure}[h]
 \includegraphics[height=9.0cm]{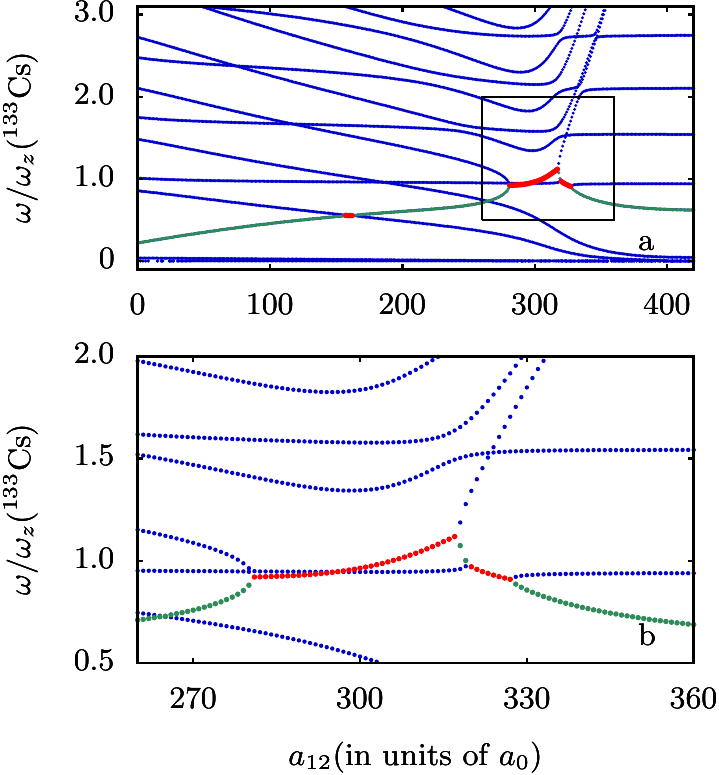}
 \caption{The evolution of the modes as a function of the interspecies
          scattering length $a_{\rm CsRb}$ in the Rb-Cs TBEC with soliton. 
          (a)The evolution of the low-lying modes  in the domain
          $0\leqslant a_{\rm CsRb}\leqslant 420 a_0 $ for $N_{\rm Rb}
          = N_{\rm Cs} = 10^3$. (b) The enlarged view of the region enclosed 
          within the black colored rectangular box in (a) to resolve the mode 
          collisions and bifurcations. The plots show only the real part of 
          mode energies $\omega/\omega_z$.         
          }
\label{mode_evol_sol}           
\end{figure}


\subsection{Interaction induced instability in TBEC}

 Dark solitons in one of the components in quasi-1D TBECs, like in 
single species, are also dynamically unstable at $T=0$ due to the quantum 
fluctuations. There is, however, another type of instability associated with 
dark solitons, and unique to TBECs. It arises from the interspecies 
interactions, and occurs when an anomalous mode collides with a higher 
energy mode. The collision transforms the two modes into degenerate complex 
energy modes, and render the dark solitonic state unstable. In the present 
work, we examine the collision of the modes as a function of the interspecies 
scattering length, and study in detail the nature of these modes, and their 
evolution. Mode collisions of similar nature, giving rise 
to {\em oscillatory unstable} states, have been investigated in the context 
of a single species cigar-shaped BEC with dark solitons in double-well 
potentials ~\cite{middelkamp_10a}.  

In TBECs, phase separation occurs when $U_{12}> \sqrt{U_{11}U_{22}}$. For
the present study, we  consider Cs and Rb as the first and second species, 
respectively. With this identification $a_{11}=a_{\rm CsCs} = 280 a_0$ and 
$a_{22}= a_{\rm RbRb}=100a_0$, and arrive at the condition for phase
separation $a_{12}=a_{_{\rm CsRb}} > 261 a_0$, which is smaller than the 
background value of $a_{_{\rm CsRb}} \approx 650a_0$ \cite{lercher_11}. To 
investigate the mode evolution with solitons, we imprint a soliton onto the 
first species (Cs condensate ) at $z=0$. We, then, vary $a_{_{\rm CsRb}}$ from 
miscible to immiscible regime, which is experimentally possible with the 
Rb-Cs Feshbach resonance \cite{pilch_09}. The mode energies, $E_j$, are
computed at $T=0$ in steps of increasing $a_{_{\rm CsRb}}$ in the domain 
$[0, 420 a_0 ]$ with $N_{\rm Rb} = N_{\rm Cs}=10^3$, 
$\omega_{z({\rm Rb})} = 2\pi\times 3.89 $Hz
and $\omega_{z({\rm Cs})} = 2\pi\times 4.55 $Hz as in
Ref. \cite{mccarron_11,pattinson_13}. To make the system quasi-1D we
take $\omega_{\perp} = 30 \omega_z$. The low-lying excitation spectrum is
characterized by the presence of an anomalous mode signifying the presence of 
soliton. The other two significant low-lying modes, which are also present 
in quasi-1D TBECs without soliton, are the Goldstone and Kohn modes of the 
two species.

When $a_{_{\rm CsRb}}=0$, the $U_{\rm CsRb}$ dependent terms in 
Eq.(\ref{bdg2}) are zero and the spectrum of the two species are independent 
as the two condensates are decoupled. The clear separation between the
modes of the two species is lost and mode mixing occurs 
when $a_{_{\rm CsRb}} > 0$. For instance, the energy of the Cs anomalous
mode increases with increasing $a_{_{\rm CsRb}}$, and collides with the other
modes resulting in the generation of a quartet of degenerate complex 
mode energies. This occurs when $a_{_{\rm CsRb}}$ is in the domains
$[157 a_0, 162 a_0]$, $[281 a_0, 317 a_0]$, and $[318 a_0, 327 a_0]$
marked by red dots in Fig.~\ref{mode_evol_sol}. In these domains, the
low-lying energy spectrum has no anomalous mode and the system is 
{\em oscillatory unstable}. For $ 162 a_0 < a_{_{\rm CsRb}}< 281 a_0$, the 
anomalous mode reappears and crosses the fourth excited state at 
$a_{_{\rm CsRb}}\approx 264 a_0$. Continuing further, as evident from 
Fig.~\ref{mode_evol_sol}(b), at $a_{_{\rm CsRb}}\approx 327 a_0$ there is a
bifurcation after which the anomalous mode ceases to undergo mode collisions. 

It should be emphasized here that, with the transition from miscible to
immiscible regime the Kohn mode and the fourth excited modes go
soft. This introduces two new Goldstone modes, including which, there are
four Goldstone modes in the excitation spectrum. These features deserve 
detailed discussion and are given in the following sections. 
\begin{figure}[ht]
 \includegraphics[width=8.5cm]{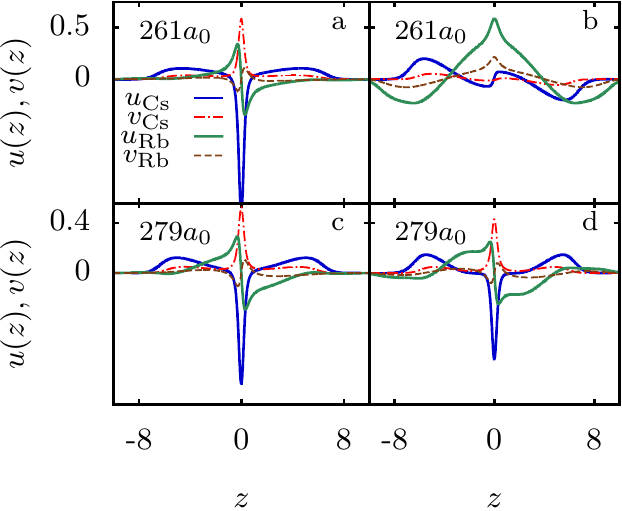}
  \caption{Variation in the nature of mode evolution near mode crossing and 
           collision. (a-b)Quasi-particle amplitudes corresponding to the 
           anomalous and fourth excited mode, respectively, at 
           $a_{_{\rm CsRb}}=261a_0$ when the modes cross each other.
           (c-d)Quasi-particle amplitudes corresponding to the anomalous and
           sixth excited mode, respectively, at $a_{_{\rm CsRb}}=279a_0$
           when the modes collide. For better visibility $u_{\rm Cs}$ and 
           $u_{\rm Rb}$ are scaled by a factor of 2.5. In the plots $u$'s and 
           $v$'s are in units of $a_{\rm osc}^{-1/2}$.
           }
\label{mode_collision_f}
\end{figure}


\subsubsection{Mode collisions}
From Fig.~\ref{mode_evol_sol}, it is evident that there are several
instances of avoided crossings and {\em mode collisions} when two modes meet
as $a_{_{\rm CsRb}}$ is varied to higher values. We have used the latter term 
(mode collision) to identify the case when one of the two modes
is the anomalous mode and when mode collisions do happen, the evolution of
the mode energies is different from the avoided crossings. In mode 
collisions, there are two possible scenarios: either the two modes cross each 
other or  undergo bifurcation. These occur due to the changes in the spatial 
profile of the mode functions ($u_{\rm Rb}$, $v_{\rm Rb}$, $u_{\rm Cs}$ and 
$v_{\rm Cs}$), which in turn depend on the condensate densities $n_{ck}(z)$. 

 To examine the case of two modes crossing each other during mode collision,
 consider the anomalous and fourth excited mode in the 
neighborhood of $a_{_{\rm CsRb}}=261a_0$. At values of $a_{_{\rm
CsRb}}$ slightly below $261a_0$, the anomalous and the fourth excited mode
approach and cross each other at $a_{_{\rm CsRb}}\approx 261a_0$. In this 
case, there are no mode mixing pre and post mode collision. 
As shown in Fig.~\ref{mode_collision_f}(a), the mode functions 
$u_{\rm Rb}$ and $v_{\rm Rb}$ corresponding to the anomalous mode are  
zero at $z=0$, whereas the mode functions $u_{\rm Cs}$
and $v_{\rm Cs}$, have maxima at $z=0$. In contrast, 
the fourth excited mode has $u_{\rm Cs}$ and $v_{\rm Cs}$ which are zero 
at $z=0$, while $u_{\rm Rb}$ and $v_{\rm Rb}$ have maxima at $z=0$ as 
shown in  Fig.~\ref{mode_collision_f}(b). 
The mode functions, thus, have very different profiles at $z=0$ and mode 
mixing does not occur, instead they just cross through.

Now, let us consider the case of bifurcation at 
$a_{_{\rm CsRb}} \approx 279a_0$. For this value of $a_{_{\rm CsRb}}$ the mode 
functions corresponding to the anomalous mode and the sixth mode have similar 
profiles with both $u_{\rm Cs}$, $v_{\rm Cs}\neq 0$ at $z=0$ as shown in 
Fig.~\ref{mode_collision_f}(c-d). These two modes collide and give rise to
complex mode energies. A similar trend is also observed at $a_{_{\rm
CsRb}}\approx 157 a_0$, when the Cs anomalous mode collides with the
Rb Kohn mode. In the domain $157 a_0\leqslant a_{_{\rm CsRb}}\leqslant
162a_0$, the profile of the Rb Kohn mode resembles the structure of
the Cs anomalous mode. So that after mode collision, they give rise to complex 
eigenfrequencies and makes the states {\em oscillatory unstable}. 

\begin{figure}[h]
 \includegraphics[width=8.5cm]{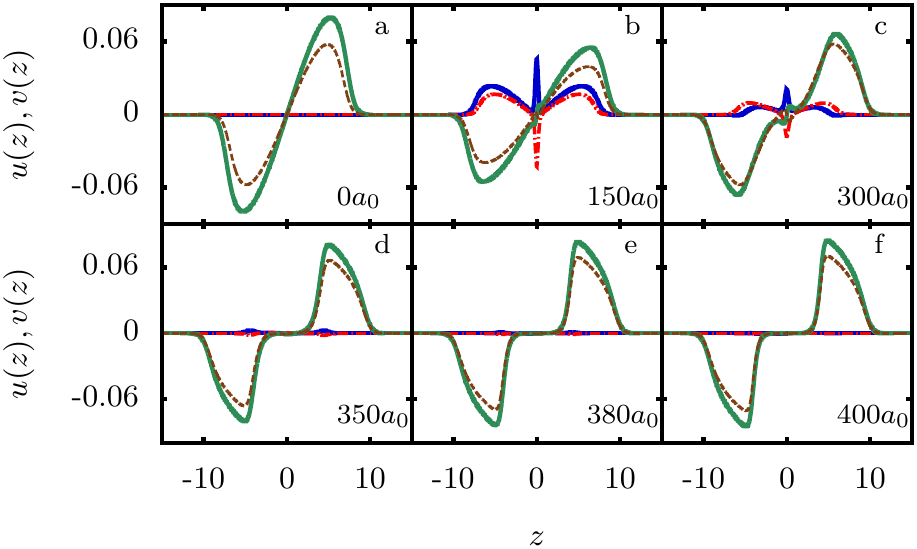}
  \caption{Evolution of quasi-particle amplitudes corresponding
          to the Rb Kohn mode as $a_{_{\rm CsRb}}$ is increased from
          0 to 400$a_0$. (a) At $a_{_{\rm CsRb}}=0$, it is a
          Kohn mode of the Rb condensate. (b-d) In the domain
          $0<a_{_{\rm CsRb}}\lesssim 350a_0$ the mode acquires admixtures from
          the Cs Kohn mode (nonzero $u_{\rm Cs}$ and $v_{\rm Cs}$). (e-f) At
          phase separation  $310a_0\lesssim a_{_{\rm CsRb}}$ the mode
          transforms to a Goldstone mode: $u_{\rm Rb}$ and $v_{\rm Rb}$
          resemble the profile of $n_{\rm Rb}=|\phi_{\rm Rb}|^2$ but with a 
          phase difference. In the plots $u$'s and $v$'s are in units of
          $a_{\rm osc}^{-1/2}$.
          }
\label{eigenfunction_1_f}
\end{figure}


\subsubsection{Third and fourth Goldstone modes}

 The third Goldstone mode emerges in the excitation spectrum as
$a_{_{\rm CsRb}}$ is increased, and the Rb Kohn mode goes soft at phase
separation when $a_{_{\rm CsRb}}\approx 350a_0$. This is consistent with 
the results reported in our earlier work ~\cite{roy_14}. The evolution of 
the Rb 
Kohn mode functions ($u_{\rm Rb}$ and $v_{\rm Rb}$) with $a_{_{\rm CsRb}}$ 
are shown in Fig. \ref{eigenfunction_1_f}. It is evident that when 
$a_{_{\rm CsRb}}=0$ (Fig. \ref{eigenfunction_1_f}(a)), there is no
admixture from the Cs Kohn mode ( $u_{\rm Cs}=v_{\rm Cs}=0$). However, when
$0<a_{_{\rm CsRb}}\lesssim 400a_0$ the admixture from the Cs Kohn mode
increases initially, and decreases to zero as we approach
$U_{\rm CsRb}> \sqrt{U_{\rm CsCs}U_{\rm RbRb}}$
(Fig.~\ref{eigenfunction_1_f}(b-f)). So, the third Goldstone mode is 
present in the system when $a_{_{\rm CsRb}}\gtrsim 350a_0$.

\begin{figure}[h]
 \includegraphics[width=8.5cm]{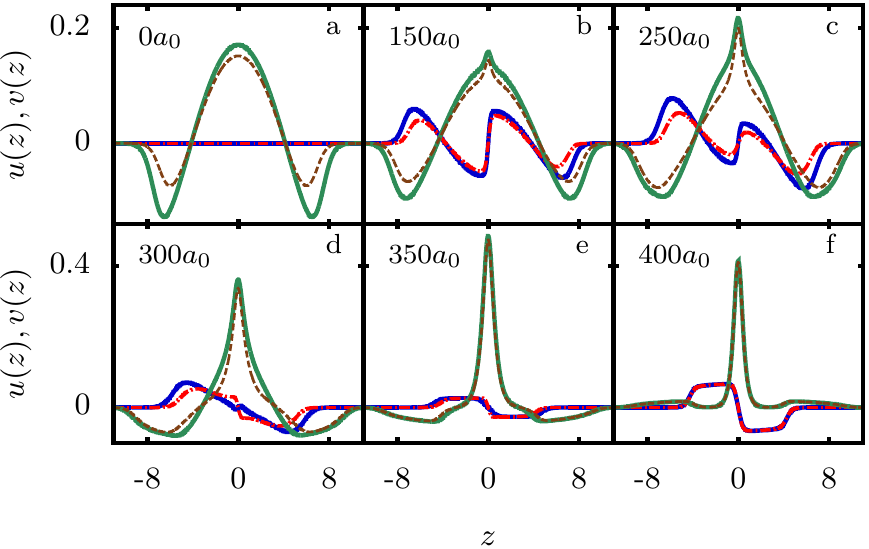}
  \caption{Evolution of the quasi-particle amplitudes corresponding
          to the fourth excited mode as $a_{_{\rm CsRb}}$ is increased from
          0 to 420$a_0$. (a) At $a_{_{\rm CsRb}}=0$, it is the
          second excited mode of the Rb condensate. (b-d) In the domain
          $0<a_{_{\rm CsRb}}\lesssim 300a_0$ the mode acquires admixtures from
          the Cs Kohn mode (nonzero $u_{\rm Cs}$ and $v_{\rm Cs}$). (e-f) At
          phase separation  $380a_0\lesssim a_{_{\rm CsRb}}$ the mode
          transforms to a Goldstone mode: $u_{\rm Rb}$, $v_{\rm Rb}$  and
          $u_{\rm Cs}$, $v_{\rm Cs}$ resemble the profile of 
          $n_{\rm Rb}=|\phi_{\rm Rb}|^2$ and
          $n_{\rm Cs}=|\phi_{\rm Cs}|^2$ but with a phase
          difference. In the plots $u$'s and $v$'s are in units of
          $a_{\rm osc}^{-1/2}$.
          }
\label{eigenfunction_2_f}
\end{figure}

The fourth excited mode, unlike in the case of quasi-1D TBECs without a
soliton also goes soft at $a_{_{\rm CsRb}}\approx 380a_0$. The evolution of
the mode functions ($u_{\rm Rb}$ and $v_{\rm Rb}$) corresponding to the 
fourth excited mode  with $a_{_{\rm CsRb}}$ are shown
in Fig. \ref{eigenfunction_2_f}. It is noticeable that when $a_{_{\rm
CsRb}}=0$ (Fig. \ref{eigenfunction_2_f}(a)), there is no contribution from
higher energy modes of Cs. However, when
$0<a_{_{\rm CsRb}}$ the admixture from the third excited mode 
of the Cs condensate is discernible in the lower values of $a_{_{\rm CsRb}}$
and are shown in Fig.~\ref{eigenfunction_2_f}(b-c). At higher values
of $a_{_{\rm CsRb}}$,  $261a_0\lesssim a_{_{\rm CsRb}}\lesssim 400a_0$, the
spatial profile of the mode functions are different from those of the
lower values of $a_{_{\rm CsRb}}$, and are shown in 
Fig.~\ref{eigenfunction_1_f}(d-f). At around $a_{_{\rm CsRb}}\approx 300a_0$, 
the mode functions begin to
resemble the structure of  $\phi_{\rm Rb}$, and the transformation is 
complete at $a_{_{\rm CsRb}}\approx 380a_0$ when the mode goes soft.

\subsection{Different mass ratios}

To gain insight on the complex nature of the mode evolution in the Rb-Cs TBEC,
we study the interplay of mass difference and intra-species scattering lengths.
For the set of aforementioned parameters the ground state of TBEC, after
phase separation is of {\em sandwich} geometry, in which the species with the 
heavier mass (Cs) is at the center and flanked by the species with lighter 
mass (Rb) at the edges ~\cite{mccarron_11}, albeit
$a_{\rm CsCs} \gg a_{\rm RbRb}$.  This geometry minimizes the trapping potential
energy, and hence the total energy of the system. In contrast, for TBECs 
with $ m_1\approx m_2$, at phase separation, the species with the smaller 
intraspecies scattering length is surrounded by the other species.
In this case the mode evolution in the presence of soliton is devoid of any 
{\em mode collisions}. Thus, we attribute the pattern of mode collisions in 
Rb-Cs TBEC binary condensate with soliton to the interplay between mass 
difference and intra-species scattering lengths.
\begin{figure}[ht]
 \includegraphics[width=8.5cm]{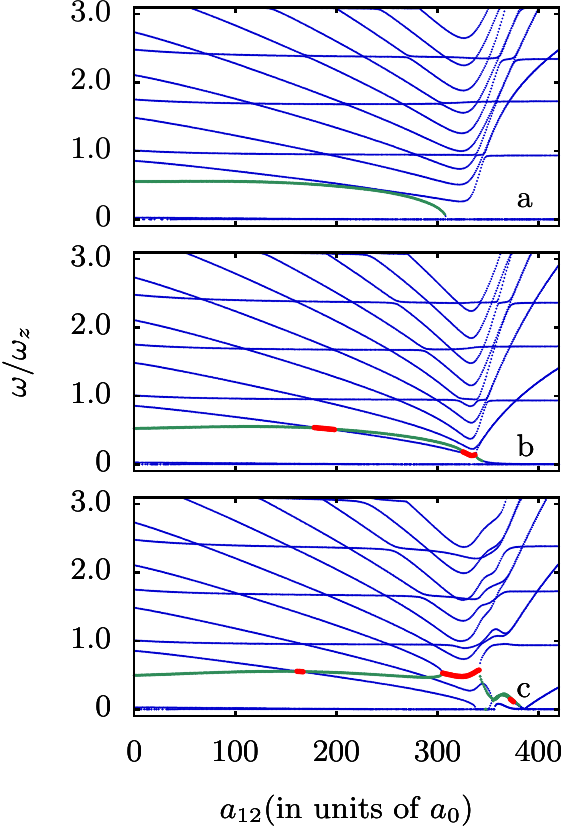}
  \caption{The evolution of the low-lying modes of the TBEC with soliton for
           different mass ratios as a function of the inter-species scattering
           length $a_{12}$ in the domain
           $0\leqslant a_{12}\leqslant 420 a_0 $. The masses of the first
           and second species in
           each of the panels correspond to (a)  95 and 87, (b) 100 and 87,
           and (c) 105 and 87 amu, respectively. The number of atoms in each
           species is $10^3$. The intra-species scattering lengths of the 
           first and second species are $a_{11}=280a_0 $  and $a_{22}=100a_0$, 
           respectively.  The plots show only the real part of mode energies 
           $\omega/\omega_z$.
           }
\label{mode_evol_soliton3combd}
\end{figure}

  To understand the transition in the mode evolution from $m_1\approx m_2$ to 
a case similar to Rb-Cs TBEC, we consider a test case where 
$87\, {\rm amu} \leqslant m_1 \leqslant 125\, {\rm amu}$  and fix 
$m_2=m_{\rm Rb}$. We then compute the evolution of the modes as a function of  
the inter-species scattering length as we increase $m_1$ from 87 amu to 
125 amu in steps of 2 amu. For example, the mode evolution for three 
different values of $m_1$ (95 amu, 100 amu, and 105 amu) are shown in 
Fig.~\ref{mode_evol_soliton3combd}. From Fig.~\ref{mode_evol_soliton3combd}(a) 
it is evident that at $m_1=95$ amu the anomalous mode goes soft
at phase separation and becomes the third Goldstone mode of the system
without any mode collisions.  At $a_{12}\approx 300a_0$, the two species are 
partially miscible and the notch of $n_1$ at $z=0$ due to the soliton is 
filled with the second species. For higher values of $a_{12}\approx 340 a_0$, 
the energetically favorable state is of a {\em sandwich geometry} where the 
species with the heavier mass ($m_1 = 95$ amu) is at the edge 
of the trap and the species with lower mass ($m_2 = 87$ amu) occupies the 
center. It should, however be recalled here that $a_{11}>a_{22}$. 

  There is a major change in the nature of mode evolution, as shown in 
Fig.~\ref{mode_evol_soliton3combd}(b) for $m_1=100$: the anomalous mode 
collides with the second excited mode twice at $a_{12}\approx 180a_0$ and 
$320a_0$. The emergence of a bifurcation is evident in the second mode 
collision at $a_{12}\approx 320a_0$.  On further increase of $m_1$, as shown in 
Fig.~\ref{mode_evol_soliton3combd}(c) for $m_1=105$, the trend of the mode 
collision begins to resemble that of the Rb-Cs mixture. 
In this case, the bifurcation arising from the collision between the anomalous 
and sixth excited mode is quite evident. Coming to the topology of the 
density profiles, prior to phase separation ( $a_{12}\approx
300a_0$) $n_1$ and $n_2$ overlap with each other and the notch of the soliton
is filled by the second species. At still higher values of $a_{12}$, $n_2$ 
from the edges migrates towards the notch of the soliton and the soliton 
gets topologically deformed. This is the energetically favorable density 
configuration. At $a_{12}\approx 380a_0$, the migration is complete 
and $n_2$ occupies the center of the trap and is surrounded by $n_1$ and the 
system is then phase-separated. Here, it must be mentioned that
without soliton the density profile would be opposite: condensates with 
masses $m_1$ and $m_2$ occupy the center and edges, respectively. Thus, the 
presence of the soliton induces a change in the topology of the density 
profiles in TBECs. On further increase of $m_1$, the energy of 
the anomalous mode increases with increasing $a_{12}$ and the collision 
with the sixth mode occurs at higher energies.


\section{Conclusions}
\label{conclusion}
In conclusion, we have examined the stability of solitons in single and
two-component BEC. We have predicted that at zero temperature presence
of soliton enhances the quantum depletion and fills up the notch of the
soliton which makes it oscillatory unstable. In TBECs having a dark soliton 
with strong interspecies interaction, four Goldstone modes emerge in the 
excitation spectrum. We have also predicted that the TBECs with soliton in 
one of the components oscillate while interacting even at zero temperature. 
This is due to the non-zero density of the other species within the notch of 
the dark soliton. We have also shown a soliton induced change in the density 
profiles when the atomic masses of the two species differ widely. Based on a 
series of computations, we find an enhancement in the mass ratio at which the 
heavier species, with higher scattering length, occupies the central position
at phase separation.


\begin{acknowledgments}
We thank K. Suthar, S. Chattopadhyay and S. Gautam for useful
discussions. The results presented in the paper are based on the
computations using the 3TFLOP HPC Cluster at Physical Research Laboratory, 
Ahmedabad, India. We also thank the anonymous referee for the thorough
review and valuable comments, which contributed to improving the quality of 
the manuscript.
\end{acknowledgments}

\bibliography{bec}{}

\begin{thebibliography}{61}%
\makeatletter
\providecommand \@ifxundefined [1]{%
 \@ifx{#1\undefined}
}%
\providecommand \@ifnum [1]{%
 \ifnum #1\expandafter \@firstoftwo
 \else \expandafter \@secondoftwo
 \fi
}%
\providecommand \@ifx [1]{%
 \ifx #1\expandafter \@firstoftwo
 \else \expandafter \@secondoftwo
 \fi
}%
\providecommand \natexlab [1]{#1}%
\providecommand \enquote  [1]{``#1''}%
\providecommand \bibnamefont  [1]{#1}%
\providecommand \bibfnamefont [1]{#1}%
\providecommand \citenamefont [1]{#1}%
\providecommand \href@noop [0]{\@secondoftwo}%
\providecommand \href [0]{\begingroup \@sanitize@url \@href}%
\providecommand \@href[1]{\@@startlink{#1}\@@href}%
\providecommand \@@href[1]{\endgroup#1\@@endlink}%
\providecommand \@sanitize@url [0]{\catcode `\\12\catcode `\$12\catcode
  `\&12\catcode `\#12\catcode `\^12\catcode `\_12\catcode `\%12\relax}%
\providecommand \@@startlink[1]{}%
\providecommand \@@endlink[0]{}%
\providecommand \url  [0]{\begingroup\@sanitize@url \@url }%
\providecommand \@url [1]{\endgroup\@href {#1}{\urlprefix }}%
\providecommand \urlprefix  [0]{URL }%
\providecommand \Eprint [0]{\href }%
\providecommand \doibase [0]{http://dx.doi.org/}%
\providecommand \selectlanguage [0]{\@gobble}%
\providecommand \bibinfo  [0]{\@secondoftwo}%
\providecommand \bibfield  [0]{\@secondoftwo}%
\providecommand \translation [1]{[#1]}%
\providecommand \BibitemOpen [0]{}%
\providecommand \bibitemStop [0]{}%
\providecommand \bibitemNoStop [0]{.\EOS\space}%
\providecommand \EOS [0]{\spacefactor3000\relax}%
\providecommand \BibitemShut  [1]{\csname bibitem#1\endcsname}%
\let\auto@bib@innerbib\@empty
\bibitem [{\citenamefont {Zurek}(1985)}]{zurek_85}%
  \BibitemOpen
  \bibfield  {author} {\bibinfo {author} {\bibfnamefont {W.~H.}\ \bibnamefont
  {Zurek}},\ }\href {\doibase 10.1038/317505a0} {\bibfield  {journal} {\bibinfo
   {journal} {Nature}\ }\textbf {\bibinfo {volume} {317}},\ \bibinfo {pages}
  {505} (\bibinfo {year} {1985})}\BibitemShut {NoStop}%
\bibitem [{\citenamefont {Zurek}(2009)}]{zurek_09}%
  \BibitemOpen
  \bibfield  {author} {\bibinfo {author} {\bibfnamefont {W.~H.}\ \bibnamefont
  {Zurek}},\ }\href {\doibase 10.1103/PhysRevLett.102.105702} {\bibfield
  {journal} {\bibinfo  {journal} {Phys. Rev. Lett.}\ }\textbf {\bibinfo
  {volume} {102}},\ \bibinfo {pages} {105702} (\bibinfo {year}
  {2009})}\BibitemShut {NoStop}%
\bibitem [{\citenamefont {Damski}\ and\ \citenamefont
  {Zurek}(2010)}]{damski_10}%
  \BibitemOpen
  \bibfield  {author} {\bibinfo {author} {\bibfnamefont {B.}~\bibnamefont
  {Damski}}\ and\ \bibinfo {author} {\bibfnamefont {W.~H.}\ \bibnamefont
  {Zurek}},\ }\href {\doibase 10.1103/PhysRevLett.104.160404} {\bibfield
  {journal} {\bibinfo  {journal} {Phys. Rev. Lett.}\ }\textbf {\bibinfo
  {volume} {104}},\ \bibinfo {pages} {160404} (\bibinfo {year}
  {2010})}\BibitemShut {NoStop}%
\bibitem [{\citenamefont {Lamporesi}\ \emph {et~al.}(2013)\citenamefont
  {Lamporesi}, \citenamefont {Donadello}, \citenamefont {Serafini},
  \citenamefont {Dalfovo},\ and\ \citenamefont {Ferrari}}]{lamporesi_13}%
  \BibitemOpen
  \bibfield  {author} {\bibinfo {author} {\bibfnamefont {G.}~\bibnamefont
  {Lamporesi}}, \bibinfo {author} {\bibfnamefont {S.}~\bibnamefont
  {Donadello}}, \bibinfo {author} {\bibfnamefont {S.}~\bibnamefont {Serafini}},
  \bibinfo {author} {\bibfnamefont {F.}~\bibnamefont {Dalfovo}}, \ and\
  \bibinfo {author} {\bibfnamefont {G.}~\bibnamefont {Ferrari}},\ }\href
  {\doibase 10.1038/nphys2734} {\bibfield  {journal} {\bibinfo  {journal} {Nat.
  Phys.}\ }\textbf {\bibinfo {volume} {9}},\ \bibinfo {pages} {656} (\bibinfo
  {year} {2013})}\BibitemShut {NoStop}%
\bibitem [{\citenamefont {Negretti}\ and\ \citenamefont
  {Henkel}(2004)}]{negretti_04}%
  \BibitemOpen
  \bibfield  {author} {\bibinfo {author} {\bibfnamefont {A.}~\bibnamefont
  {Negretti}}\ and\ \bibinfo {author} {\bibfnamefont {C.}~\bibnamefont
  {Henkel}},\ }\href {http://stacks.iop.org/0953-4075/37/i=23/a=L02} {\bibfield
   {journal} {\bibinfo  {journal} {J. Phys. B}\ }\textbf {\bibinfo {volume}
  {37}},\ \bibinfo {pages} {L385} (\bibinfo {year} {2004})}\BibitemShut
  {NoStop}%
\bibitem [{\citenamefont {Negretti}\ \emph {et~al.}(2008)\citenamefont
  {Negretti}, \citenamefont {Henkel},\ and\ \citenamefont
  {M\o{}lmer}}]{negretti_08}%
  \BibitemOpen
  \bibfield  {author} {\bibinfo {author} {\bibfnamefont {A.}~\bibnamefont
  {Negretti}}, \bibinfo {author} {\bibfnamefont {C.}~\bibnamefont {Henkel}}, \
  and\ \bibinfo {author} {\bibfnamefont {K.}~\bibnamefont {M\o{}lmer}},\ }\href
  {\doibase 10.1103/PhysRevA.78.023630} {\bibfield  {journal} {\bibinfo
  {journal} {Phys. Rev. A}\ }\textbf {\bibinfo {volume} {78}},\ \bibinfo
  {pages} {023630} (\bibinfo {year} {2008})}\BibitemShut {NoStop}%
\bibitem [{\citenamefont {Burger}\ \emph {et~al.}(1999)\citenamefont {Burger},
  \citenamefont {Bongs}, \citenamefont {Dettmer}, \citenamefont {Ertmer},
  \citenamefont {Sengstock}, \citenamefont {Sanpera}, \citenamefont
  {Shlyapnikov},\ and\ \citenamefont {Lewenstein}}]{burger_99}%
  \BibitemOpen
  \bibfield  {author} {\bibinfo {author} {\bibfnamefont {S.}~\bibnamefont
  {Burger}}, \bibinfo {author} {\bibfnamefont {K.}~\bibnamefont {Bongs}},
  \bibinfo {author} {\bibfnamefont {S.}~\bibnamefont {Dettmer}}, \bibinfo
  {author} {\bibfnamefont {W.}~\bibnamefont {Ertmer}}, \bibinfo {author}
  {\bibfnamefont {K.}~\bibnamefont {Sengstock}}, \bibinfo {author}
  {\bibfnamefont {A.}~\bibnamefont {Sanpera}}, \bibinfo {author} {\bibfnamefont
  {G.~V.}\ \bibnamefont {Shlyapnikov}}, \ and\ \bibinfo {author} {\bibfnamefont
  {M.}~\bibnamefont {Lewenstein}},\ }\href {\doibase
  10.1103/PhysRevLett.83.5198} {\bibfield  {journal} {\bibinfo  {journal}
  {Phys. Rev. Lett.}\ }\textbf {\bibinfo {volume} {83}},\ \bibinfo {pages}
  {5198} (\bibinfo {year} {1999})}\BibitemShut {NoStop}%
\bibitem [{\citenamefont {Denschlag}\ \emph {et~al.}(2000)\citenamefont
  {Denschlag}, \citenamefont {Simsarian}, \citenamefont {Feder}, \citenamefont
  {Clark}, \citenamefont {Collins}, \citenamefont {Cubizolles}, \citenamefont
  {Deng}, \citenamefont {Hagley}, \citenamefont {Helmerson}, \citenamefont
  {Reinhardt}, \citenamefont {Rolston}, \citenamefont {Schneider},\ and\
  \citenamefont {Phillips}}]{denschlag_00}%
  \BibitemOpen
  \bibfield  {author} {\bibinfo {author} {\bibfnamefont {J.}~\bibnamefont
  {Denschlag}}, \bibinfo {author} {\bibfnamefont {J.~E.}\ \bibnamefont
  {Simsarian}}, \bibinfo {author} {\bibfnamefont {D.~L.}\ \bibnamefont
  {Feder}}, \bibinfo {author} {\bibfnamefont {C.~W.}\ \bibnamefont {Clark}},
  \bibinfo {author} {\bibfnamefont {L.~A.}\ \bibnamefont {Collins}}, \bibinfo
  {author} {\bibfnamefont {J.}~\bibnamefont {Cubizolles}}, \bibinfo {author}
  {\bibfnamefont {L.}~\bibnamefont {Deng}}, \bibinfo {author} {\bibfnamefont
  {E.~W.}\ \bibnamefont {Hagley}}, \bibinfo {author} {\bibfnamefont
  {K.}~\bibnamefont {Helmerson}}, \bibinfo {author} {\bibfnamefont {W.~P.}\
  \bibnamefont {Reinhardt}}, \bibinfo {author} {\bibfnamefont {S.~L.}\
  \bibnamefont {Rolston}}, \bibinfo {author} {\bibfnamefont {B.~I.}\
  \bibnamefont {Schneider}}, \ and\ \bibinfo {author} {\bibfnamefont {W.~D.}\
  \bibnamefont {Phillips}},\ }\href {\doibase 10.1126/science.287.5450.97}
  {\bibfield  {journal} {\bibinfo  {journal} {Science}\ }\textbf {\bibinfo
  {volume} {287}},\ \bibinfo {pages} {97} (\bibinfo {year} {2000})}\BibitemShut
  {NoStop}%
\bibitem [{\citenamefont {Busch}\ and\ \citenamefont
  {Anglin}(2000)}]{busch_00}%
  \BibitemOpen
  \bibfield  {author} {\bibinfo {author} {\bibfnamefont {T.}~\bibnamefont
  {Busch}}\ and\ \bibinfo {author} {\bibfnamefont {J.~R.}\ \bibnamefont
  {Anglin}},\ }\href {\doibase 10.1103/PhysRevLett.84.2298} {\bibfield
  {journal} {\bibinfo  {journal} {Phys. Rev. Lett.}\ }\textbf {\bibinfo
  {volume} {84}},\ \bibinfo {pages} {2298} (\bibinfo {year}
  {2000})}\BibitemShut {NoStop}%
\bibitem [{\citenamefont {Busch}\ and\ \citenamefont
  {Anglin}(2001)}]{busch_01}%
  \BibitemOpen
  \bibfield  {author} {\bibinfo {author} {\bibfnamefont {T.}~\bibnamefont
  {Busch}}\ and\ \bibinfo {author} {\bibfnamefont {J.~R.}\ \bibnamefont
  {Anglin}},\ }\href {\doibase 10.1103/PhysRevLett.87.010401} {\bibfield
  {journal} {\bibinfo  {journal} {Phys. Rev. Lett.}\ }\textbf {\bibinfo
  {volume} {87}},\ \bibinfo {pages} {010401} (\bibinfo {year}
  {2001})}\BibitemShut {NoStop}%
\bibitem [{\citenamefont {Middelkamp}\ \emph {et~al.}(2011)\citenamefont
  {Middelkamp}, \citenamefont {Chang}, \citenamefont {Hamner}, \citenamefont
  {Carretero-Gonz\'alez}, \citenamefont {Kevrekidis}, \citenamefont
  {Achilleos}, \citenamefont {Frantzeskakis}, \citenamefont {Schmelcher},\ and\
  \citenamefont {Engels}}]{middelkamp-11}%
  \BibitemOpen
  \bibfield  {author} {\bibinfo {author} {\bibfnamefont {S.}~\bibnamefont
  {Middelkamp}}, \bibinfo {author} {\bibfnamefont {J.}~\bibnamefont {Chang}},
  \bibinfo {author} {\bibfnamefont {C.}~\bibnamefont {Hamner}}, \bibinfo
  {author} {\bibfnamefont {R.}~\bibnamefont {Carretero-Gonz\'alez}}, \bibinfo
  {author} {\bibfnamefont {P.}~\bibnamefont {Kevrekidis}}, \bibinfo {author}
  {\bibfnamefont {V.}~\bibnamefont {Achilleos}}, \bibinfo {author}
  {\bibfnamefont {D.}~\bibnamefont {Frantzeskakis}}, \bibinfo {author}
  {\bibfnamefont {P.}~\bibnamefont {Schmelcher}}, \ and\ \bibinfo {author}
  {\bibfnamefont {P.}~\bibnamefont {Engels}},\ }\href {\doibase
  http://dx.doi.org/10.1016/j.physleta.2010.11.025} {\bibfield  {journal}
  {\bibinfo  {journal} {Phys. Lett. A}\ }\textbf {\bibinfo {volume} {375}},\
  \bibinfo {pages} {642 } (\bibinfo {year} {2011})}\BibitemShut {NoStop}%
\bibitem [{\citenamefont {Hoefer}\ \emph {et~al.}(2011)\citenamefont {Hoefer},
  \citenamefont {Chang}, \citenamefont {Hamner},\ and\ \citenamefont
  {Engels}}]{hoefer_11}%
  \BibitemOpen
  \bibfield  {author} {\bibinfo {author} {\bibfnamefont {M.~A.}\ \bibnamefont
  {Hoefer}}, \bibinfo {author} {\bibfnamefont {J.~J.}\ \bibnamefont {Chang}},
  \bibinfo {author} {\bibfnamefont {C.}~\bibnamefont {Hamner}}, \ and\ \bibinfo
  {author} {\bibfnamefont {P.}~\bibnamefont {Engels}},\ }\href {\doibase
  10.1103/PhysRevA.84.041605} {\bibfield  {journal} {\bibinfo  {journal} {Phys.
  Rev. A}\ }\textbf {\bibinfo {volume} {84}},\ \bibinfo {pages} {041605}
  (\bibinfo {year} {2011})}\BibitemShut {NoStop}%
\bibitem [{\citenamefont {Kasamatsu}\ and\ \citenamefont
  {Tsubota}(2006)}]{kasamatsu_06}%
  \BibitemOpen
  \bibfield  {author} {\bibinfo {author} {\bibfnamefont {K.}~\bibnamefont
  {Kasamatsu}}\ and\ \bibinfo {author} {\bibfnamefont {M.}~\bibnamefont
  {Tsubota}},\ }\href {\doibase 10.1103/PhysRevA.74.013617} {\bibfield
  {journal} {\bibinfo  {journal} {Phys. Rev. A}\ }\textbf {\bibinfo {volume}
  {74}},\ \bibinfo {pages} {013617} (\bibinfo {year} {2006})}\BibitemShut
  {NoStop}%
\bibitem [{\citenamefont {Rajendran}\ \emph {et~al.}(2009)\citenamefont
  {Rajendran}, \citenamefont {Muruganandam},\ and\ \citenamefont
  {Lakshmanan}}]{rajendran_09}%
  \BibitemOpen
  \bibfield  {author} {\bibinfo {author} {\bibfnamefont {S.}~\bibnamefont
  {Rajendran}}, \bibinfo {author} {\bibfnamefont {P.}~\bibnamefont
  {Muruganandam}}, \ and\ \bibinfo {author} {\bibfnamefont {M.}~\bibnamefont
  {Lakshmanan}},\ }\href {http://stacks.iop.org/0953-4075/42/i=14/a=145307}
  {\bibfield  {journal} {\bibinfo  {journal} {J. Phys. B}\ }\textbf {\bibinfo
  {volume} {42}},\ \bibinfo {pages} {145307} (\bibinfo {year}
  {2009})}\BibitemShut {NoStop}%
\bibitem [{\citenamefont {Achilleos}\ \emph {et~al.}(2012)\citenamefont
  {Achilleos}, \citenamefont {Yan}, \citenamefont {Kevrekidis},\ and\
  \citenamefont {Frantzeskakis}}]{achilleos_12}%
  \BibitemOpen
  \bibfield  {author} {\bibinfo {author} {\bibfnamefont {V.}~\bibnamefont
  {Achilleos}}, \bibinfo {author} {\bibfnamefont {D.}~\bibnamefont {Yan}},
  \bibinfo {author} {\bibfnamefont {P.~G.}\ \bibnamefont {Kevrekidis}}, \ and\
  \bibinfo {author} {\bibfnamefont {D.~J.}\ \bibnamefont {Frantzeskakis}},\
  }\href {http://stacks.iop.org/1367-2630/14/i=5/a=055006} {\bibfield
  {journal} {\bibinfo  {journal} {New Journal of Physics}\ }\textbf {\bibinfo
  {volume} {14}},\ \bibinfo {pages} {055006} (\bibinfo {year}
  {2012})}\BibitemShut {NoStop}%
\bibitem [{\citenamefont {Becker}\ \emph {et~al.}(2008)\citenamefont {Becker},
  \citenamefont {Stellmer}, \citenamefont {Soltan-Panahi}, \citenamefont
  {Dorscher}, \citenamefont {Baumert}, \citenamefont {Richter}, \citenamefont
  {Kronjager}, \citenamefont {Bongs},\ and\ \citenamefont
  {Sengstock}}]{becker_08}%
  \BibitemOpen
  \bibfield  {author} {\bibinfo {author} {\bibfnamefont {C.}~\bibnamefont
  {Becker}}, \bibinfo {author} {\bibfnamefont {S.}~\bibnamefont {Stellmer}},
  \bibinfo {author} {\bibfnamefont {P.}~\bibnamefont {Soltan-Panahi}}, \bibinfo
  {author} {\bibfnamefont {S.}~\bibnamefont {Dorscher}}, \bibinfo {author}
  {\bibfnamefont {M.}~\bibnamefont {Baumert}}, \bibinfo {author} {\bibfnamefont
  {E.-M.}\ \bibnamefont {Richter}}, \bibinfo {author} {\bibfnamefont
  {J.}~\bibnamefont {Kronjager}}, \bibinfo {author} {\bibfnamefont
  {K.}~\bibnamefont {Bongs}}, \ and\ \bibinfo {author} {\bibfnamefont
  {K.}~\bibnamefont {Sengstock}},\ }\href {\doibase 10.1038/nphys962}
  {\bibfield  {journal} {\bibinfo  {journal} {Nat. Phys.}\ }\textbf {\bibinfo
  {volume} {4}},\ \bibinfo {pages} {496} (\bibinfo {year} {2008})}\BibitemShut
  {NoStop}%
\bibitem [{\citenamefont {Weller}\ \emph {et~al.}(2008)\citenamefont {Weller},
  \citenamefont {Ronzheimer}, \citenamefont {Gross}, \citenamefont {Esteve},
  \citenamefont {Oberthaler}, \citenamefont {Frantzeskakis}, \citenamefont
  {Theocharis},\ and\ \citenamefont {Kevrekidis}}]{weller_08}%
  \BibitemOpen
  \bibfield  {author} {\bibinfo {author} {\bibfnamefont {A.}~\bibnamefont
  {Weller}}, \bibinfo {author} {\bibfnamefont {J.~P.}\ \bibnamefont
  {Ronzheimer}}, \bibinfo {author} {\bibfnamefont {C.}~\bibnamefont {Gross}},
  \bibinfo {author} {\bibfnamefont {J.}~\bibnamefont {Esteve}}, \bibinfo
  {author} {\bibfnamefont {M.~K.}\ \bibnamefont {Oberthaler}}, \bibinfo
  {author} {\bibfnamefont {D.~J.}\ \bibnamefont {Frantzeskakis}}, \bibinfo
  {author} {\bibfnamefont {G.}~\bibnamefont {Theocharis}}, \ and\ \bibinfo
  {author} {\bibfnamefont {P.~G.}\ \bibnamefont {Kevrekidis}},\ }\href
  {\doibase 10.1103/PhysRevLett.101.130401} {\bibfield  {journal} {\bibinfo
  {journal} {Phys. Rev. Lett.}\ }\textbf {\bibinfo {volume} {101}},\ \bibinfo
  {pages} {130401} (\bibinfo {year} {2008})}\BibitemShut {NoStop}%
\bibitem [{\citenamefont {Stellmer}\ \emph {et~al.}(2008)\citenamefont
  {Stellmer}, \citenamefont {Becker}, \citenamefont {Soltan-Panahi},
  \citenamefont {Richter}, \citenamefont {D\"orscher}, \citenamefont {Baumert},
  \citenamefont {Kronj\"ager}, \citenamefont {Bongs},\ and\ \citenamefont
  {Sengstock}}]{stellmer_08}%
  \BibitemOpen
  \bibfield  {author} {\bibinfo {author} {\bibfnamefont {S.}~\bibnamefont
  {Stellmer}}, \bibinfo {author} {\bibfnamefont {C.}~\bibnamefont {Becker}},
  \bibinfo {author} {\bibfnamefont {P.}~\bibnamefont {Soltan-Panahi}}, \bibinfo
  {author} {\bibfnamefont {E.-M.}\ \bibnamefont {Richter}}, \bibinfo {author}
  {\bibfnamefont {S.}~\bibnamefont {D\"orscher}}, \bibinfo {author}
  {\bibfnamefont {M.}~\bibnamefont {Baumert}}, \bibinfo {author} {\bibfnamefont
  {J.}~\bibnamefont {Kronj\"ager}}, \bibinfo {author} {\bibfnamefont
  {K.}~\bibnamefont {Bongs}}, \ and\ \bibinfo {author} {\bibfnamefont
  {K.}~\bibnamefont {Sengstock}},\ }\href {\doibase
  10.1103/PhysRevLett.101.120406} {\bibfield  {journal} {\bibinfo  {journal}
  {Phys. Rev. Lett.}\ }\textbf {\bibinfo {volume} {101}},\ \bibinfo {pages}
  {120406} (\bibinfo {year} {2008})}\BibitemShut {NoStop}%
\bibitem [{\citenamefont {Muryshev}\ \emph {et~al.}(1999)\citenamefont
  {Muryshev}, \citenamefont {van Linden van~den Heuvell},\ and\ \citenamefont
  {Shlyapnikov}}]{muryshev_99}%
  \BibitemOpen
  \bibfield  {author} {\bibinfo {author} {\bibfnamefont {A.~E.}\ \bibnamefont
  {Muryshev}}, \bibinfo {author} {\bibfnamefont {H.~B.}\ \bibnamefont {van
  Linden van~den Heuvell}}, \ and\ \bibinfo {author} {\bibfnamefont {G.~V.}\
  \bibnamefont {Shlyapnikov}},\ }\href {\doibase 10.1103/PhysRevA.60.R2665}
  {\bibfield  {journal} {\bibinfo  {journal} {Phys. Rev. A}\ }\textbf {\bibinfo
  {volume} {60}},\ \bibinfo {pages} {R2665} (\bibinfo {year}
  {1999})}\BibitemShut {NoStop}%
\bibitem [{\citenamefont {Middelkamp}\ \emph {et~al.}(2010)\citenamefont
  {Middelkamp}, \citenamefont {Theocharis}, \citenamefont {Kevrekidis},
  \citenamefont {Frantzeskakis},\ and\ \citenamefont
  {Schmelcher}}]{middelkamp_10a}%
  \BibitemOpen
  \bibfield  {author} {\bibinfo {author} {\bibfnamefont {S.}~\bibnamefont
  {Middelkamp}}, \bibinfo {author} {\bibfnamefont {G.}~\bibnamefont
  {Theocharis}}, \bibinfo {author} {\bibfnamefont {P.~G.}\ \bibnamefont
  {Kevrekidis}}, \bibinfo {author} {\bibfnamefont {D.~J.}\ \bibnamefont
  {Frantzeskakis}}, \ and\ \bibinfo {author} {\bibfnamefont {P.}~\bibnamefont
  {Schmelcher}},\ }\href {\doibase 10.1103/PhysRevA.81.053618} {\bibfield
  {journal} {\bibinfo  {journal} {Phys. Rev. A}\ }\textbf {\bibinfo {volume}
  {81}},\ \bibinfo {pages} {053618} (\bibinfo {year} {2010})}\BibitemShut
  {NoStop}%
\bibitem [{\citenamefont {Mochol}\ \emph {et~al.}(2012)\citenamefont {Mochol},
  \citenamefont {P\l{}odzie\ifmmode~\acute{n}\else \'{n}\fi{}},\ and\
  \citenamefont {Sacha}}]{mochol_12}%
  \BibitemOpen
  \bibfield  {author} {\bibinfo {author} {\bibfnamefont {M.}~\bibnamefont
  {Mochol}}, \bibinfo {author} {\bibfnamefont {M.}~\bibnamefont
  {P\l{}odzie\ifmmode~\acute{n}\else \'{n}\fi{}}}, \ and\ \bibinfo {author}
  {\bibfnamefont {K.}~\bibnamefont {Sacha}},\ }\href {\doibase
  10.1103/PhysRevA.85.023627} {\bibfield  {journal} {\bibinfo  {journal} {Phys.
  Rev. A}\ }\textbf {\bibinfo {volume} {85}},\ \bibinfo {pages} {023627}
  (\bibinfo {year} {2012})}\BibitemShut {NoStop}%
\bibitem [{\citenamefont {Kevrekidis}\ \emph {et~al.}(2003)\citenamefont
  {Kevrekidis}, \citenamefont {Carretero-Gonz\'alez}, \citenamefont
  {Theocharis}, \citenamefont {Frantzeskakis},\ and\ \citenamefont
  {Malomed}}]{kevrekidis_03}%
  \BibitemOpen
  \bibfield  {author} {\bibinfo {author} {\bibfnamefont {P.~G.}\ \bibnamefont
  {Kevrekidis}}, \bibinfo {author} {\bibfnamefont {R.}~\bibnamefont
  {Carretero-Gonz\'alez}}, \bibinfo {author} {\bibfnamefont {G.}~\bibnamefont
  {Theocharis}}, \bibinfo {author} {\bibfnamefont {D.~J.}\ \bibnamefont
  {Frantzeskakis}}, \ and\ \bibinfo {author} {\bibfnamefont {B.~A.}\
  \bibnamefont {Malomed}},\ }\href {\doibase 10.1103/PhysRevA.68.035602}
  {\bibfield  {journal} {\bibinfo  {journal} {Phys. Rev. A}\ }\textbf {\bibinfo
  {volume} {68}},\ \bibinfo {pages} {035602} (\bibinfo {year}
  {2003})}\BibitemShut {NoStop}%
\bibitem [{\citenamefont {Parker}\ \emph {et~al.}(2004)\citenamefont {Parker},
  \citenamefont {Proukakis}, \citenamefont {Barenghi},\ and\ \citenamefont
  {Adams}}]{parker_04}%
  \BibitemOpen
  \bibfield  {author} {\bibinfo {author} {\bibfnamefont {N.~G.}\ \bibnamefont
  {Parker}}, \bibinfo {author} {\bibfnamefont {N.~P.}\ \bibnamefont
  {Proukakis}}, \bibinfo {author} {\bibfnamefont {C.~F.}\ \bibnamefont
  {Barenghi}}, \ and\ \bibinfo {author} {\bibfnamefont {C.~S.}\ \bibnamefont
  {Adams}},\ }\href {http://stacks.iop.org/0953-4075/37/i=7/a=063} {\bibfield
  {journal} {\bibinfo  {journal} {J. Phys. B}\ }\textbf {\bibinfo {volume}
  {37}},\ \bibinfo {pages} {S175} (\bibinfo {year} {2004})}\BibitemShut
  {NoStop}%
\bibitem [{\citenamefont {Sakaguchi}\ and\ \citenamefont
  {Malomed}(2005)}]{sakaguchi_05}%
  \BibitemOpen
  \bibfield  {author} {\bibinfo {author} {\bibfnamefont {H.}~\bibnamefont
  {Sakaguchi}}\ and\ \bibinfo {author} {\bibfnamefont {B.~A.}\ \bibnamefont
  {Malomed}},\ }\href {\doibase 10.1103/PhysRevE.72.046610} {\bibfield
  {journal} {\bibinfo  {journal} {Phys. Rev. E}\ }\textbf {\bibinfo {volume}
  {72}},\ \bibinfo {pages} {046610} (\bibinfo {year} {2005})}\BibitemShut
  {NoStop}%
\bibitem [{\citenamefont {Kartashov}\ \emph {et~al.}(2011)\citenamefont
  {Kartashov}, \citenamefont {Malomed},\ and\ \citenamefont
  {Torner}}]{kartashov_11}%
  \BibitemOpen
  \bibfield  {author} {\bibinfo {author} {\bibfnamefont {Y.~V.}\ \bibnamefont
  {Kartashov}}, \bibinfo {author} {\bibfnamefont {B.~A.}\ \bibnamefont
  {Malomed}}, \ and\ \bibinfo {author} {\bibfnamefont {L.}~\bibnamefont
  {Torner}},\ }\href {\doibase 10.1103/RevModPhys.83.247} {\bibfield  {journal}
  {\bibinfo  {journal} {Rev. Mod. Phys.}\ }\textbf {\bibinfo {volume} {83}},\
  \bibinfo {pages} {247} (\bibinfo {year} {2011})}\BibitemShut {NoStop}%
\bibitem [{\citenamefont {Atre}\ \emph {et~al.}(2006)\citenamefont {Atre},
  \citenamefont {Panigrahi},\ and\ \citenamefont {Agarwal}}]{atre_06}%
  \BibitemOpen
  \bibfield  {author} {\bibinfo {author} {\bibfnamefont {R.}~\bibnamefont
  {Atre}}, \bibinfo {author} {\bibfnamefont {P.~K.}\ \bibnamefont {Panigrahi}},
  \ and\ \bibinfo {author} {\bibfnamefont {G.~S.}\ \bibnamefont {Agarwal}},\
  }\href {\doibase 10.1103/PhysRevE.73.056611} {\bibfield  {journal} {\bibinfo
  {journal} {Phys. Rev. E}\ }\textbf {\bibinfo {volume} {73}},\ \bibinfo
  {pages} {056611} (\bibinfo {year} {2006})}\BibitemShut {NoStop}%
\bibitem [{\citenamefont {Theocharis}\ \emph {et~al.}(2010)\citenamefont
  {Theocharis}, \citenamefont {Weller}, \citenamefont {Ronzheimer},
  \citenamefont {Gross}, \citenamefont {Oberthaler}, \citenamefont
  {Kevrekidis},\ and\ \citenamefont {Frantzeskakis}}]{theocharis_10}%
  \BibitemOpen
  \bibfield  {author} {\bibinfo {author} {\bibfnamefont {G.}~\bibnamefont
  {Theocharis}}, \bibinfo {author} {\bibfnamefont {A.}~\bibnamefont {Weller}},
  \bibinfo {author} {\bibfnamefont {J.~P.}\ \bibnamefont {Ronzheimer}},
  \bibinfo {author} {\bibfnamefont {C.}~\bibnamefont {Gross}}, \bibinfo
  {author} {\bibfnamefont {M.~K.}\ \bibnamefont {Oberthaler}}, \bibinfo
  {author} {\bibfnamefont {P.~G.}\ \bibnamefont {Kevrekidis}}, \ and\ \bibinfo
  {author} {\bibfnamefont {D.~J.}\ \bibnamefont {Frantzeskakis}},\ }\href
  {\doibase 10.1103/PhysRevA.81.063604} {\bibfield  {journal} {\bibinfo
  {journal} {Phys. Rev. A}\ }\textbf {\bibinfo {volume} {81}},\ \bibinfo
  {pages} {063604} (\bibinfo {year} {2010})}\BibitemShut {NoStop}%
\bibitem [{\citenamefont {Dziarmaga}\ and\ \citenamefont
  {Sacha}(2002)}]{dziarmaga-02}%
  \BibitemOpen
  \bibfield  {author} {\bibinfo {author} {\bibfnamefont {J.}~\bibnamefont
  {Dziarmaga}}\ and\ \bibinfo {author} {\bibfnamefont {K.}~\bibnamefont
  {Sacha}},\ }\href {\doibase 10.1103/PhysRevA.66.043620} {\bibfield  {journal}
  {\bibinfo  {journal} {Phys. Rev. A}\ }\textbf {\bibinfo {volume} {66}},\
  \bibinfo {pages} {043620} (\bibinfo {year} {2002})}\BibitemShut {NoStop}%
\bibitem [{\citenamefont {Dziarmaga}\ \emph {et~al.}(2002)\citenamefont
  {Dziarmaga}, \citenamefont {Karkuszewski},\ and\ \citenamefont
  {Sacha}}]{dziarmaga_02}%
  \BibitemOpen
  \bibfield  {author} {\bibinfo {author} {\bibfnamefont {J.}~\bibnamefont
  {Dziarmaga}}, \bibinfo {author} {\bibfnamefont {Z.~P.}\ \bibnamefont
  {Karkuszewski}}, \ and\ \bibinfo {author} {\bibfnamefont {K.}~\bibnamefont
  {Sacha}},\ }\href {\doibase 10.1103/PhysRevA.66.043615} {\bibfield  {journal}
  {\bibinfo  {journal} {Phys. Rev. A}\ }\textbf {\bibinfo {volume} {66}},\
  \bibinfo {pages} {043615} (\bibinfo {year} {2002})}\BibitemShut {NoStop}%
\bibitem [{\citenamefont {Dziarmaga}\ and\ \citenamefont
  {Sacha}(2003)}]{dziarmaga_03}%
  \BibitemOpen
  \bibfield  {author} {\bibinfo {author} {\bibfnamefont {J.}~\bibnamefont
  {Dziarmaga}}\ and\ \bibinfo {author} {\bibfnamefont {K.}~\bibnamefont
  {Sacha}},\ }\href {\doibase 10.1103/PhysRevA.67.033608} {\bibfield  {journal}
  {\bibinfo  {journal} {Phys. Rev. A}\ }\textbf {\bibinfo {volume} {67}},\
  \bibinfo {pages} {033608} (\bibinfo {year} {2003})}\BibitemShut {NoStop}%
\bibitem [{\citenamefont {Dziarmaga}\ \emph {et~al.}(2003)\citenamefont
  {Dziarmaga}, \citenamefont {Karkuszewski},\ and\ \citenamefont
  {Sacha}}]{dziarmaga-03}%
  \BibitemOpen
  \bibfield  {author} {\bibinfo {author} {\bibfnamefont {J.}~\bibnamefont
  {Dziarmaga}}, \bibinfo {author} {\bibfnamefont {Z.~P.}\ \bibnamefont
  {Karkuszewski}}, \ and\ \bibinfo {author} {\bibfnamefont {K.}~\bibnamefont
  {Sacha}},\ }\href {http://stacks.iop.org/0953-4075/36/i=6/a=311} {\bibfield
  {journal} {\bibinfo  {journal} {J. Phys. B}\ }\textbf {\bibinfo {volume}
  {36}},\ \bibinfo {pages} {1217} (\bibinfo {year} {2003})}\BibitemShut
  {NoStop}%
\bibitem [{\citenamefont {Law}\ \emph {et~al.}(2002)\citenamefont {Law},
  \citenamefont {Leung},\ and\ \citenamefont {Chu}}]{law_02}%
  \BibitemOpen
  \bibfield  {author} {\bibinfo {author} {\bibfnamefont {C.~K.}\ \bibnamefont
  {Law}}, \bibinfo {author} {\bibfnamefont {P.~T.}\ \bibnamefont {Leung}}, \
  and\ \bibinfo {author} {\bibfnamefont {M.-C.}\ \bibnamefont {Chu}},\ }\href
  {http://stacks.iop.org/0953-4075/35/i=16/a=316} {\bibfield  {journal}
  {\bibinfo  {journal} {J. Phys. B}\ }\textbf {\bibinfo {volume} {35}},\
  \bibinfo {pages} {3583} (\bibinfo {year} {2002})}\BibitemShut {NoStop}%
\bibitem [{\citenamefont {Law}(2003)}]{law_03}%
  \BibitemOpen
  \bibfield  {author} {\bibinfo {author} {\bibfnamefont {C.~K.}\ \bibnamefont
  {Law}},\ }\href {\doibase 10.1103/PhysRevA.68.015602} {\bibfield  {journal}
  {\bibinfo  {journal} {Phys. Rev. A}\ }\textbf {\bibinfo {volume} {68}},\
  \bibinfo {pages} {015602} (\bibinfo {year} {2003})}\BibitemShut {NoStop}%
\bibitem [{\citenamefont {Gangardt}\ and\ \citenamefont
  {Kamenev}(2010)}]{gangardt_10}%
  \BibitemOpen
  \bibfield  {author} {\bibinfo {author} {\bibfnamefont {D.~M.}\ \bibnamefont
  {Gangardt}}\ and\ \bibinfo {author} {\bibfnamefont {A.}~\bibnamefont
  {Kamenev}},\ }\href {\doibase 10.1103/PhysRevLett.104.190402} {\bibfield
  {journal} {\bibinfo  {journal} {Phys. Rev. Lett.}\ }\textbf {\bibinfo
  {volume} {104}},\ \bibinfo {pages} {190402} (\bibinfo {year}
  {2010})}\BibitemShut {NoStop}%
\bibitem [{\citenamefont {Papp}\ \emph {et~al.}(2008)\citenamefont {Papp},
  \citenamefont {Pino},\ and\ \citenamefont {Wieman}}]{papp_08}%
  \BibitemOpen
  \bibfield  {author} {\bibinfo {author} {\bibfnamefont {S.~B.}\ \bibnamefont
  {Papp}}, \bibinfo {author} {\bibfnamefont {J.~M.}\ \bibnamefont {Pino}}, \
  and\ \bibinfo {author} {\bibfnamefont {C.~E.}\ \bibnamefont {Wieman}},\
  }\href {\doibase 10.1103/PhysRevLett.101.040402} {\bibfield  {journal}
  {\bibinfo  {journal} {Phys. Rev. Lett.}\ }\textbf {\bibinfo {volume} {101}},\
  \bibinfo {pages} {040402} (\bibinfo {year} {2008})}\BibitemShut {NoStop}%
\bibitem [{\citenamefont {Tojo}\ \emph {et~al.}(2010)\citenamefont {Tojo},
  \citenamefont {Taguchi}, \citenamefont {Masuyama}, \citenamefont {Hayashi},
  \citenamefont {Saito},\ and\ \citenamefont {Hirano}}]{tojo_10}%
  \BibitemOpen
  \bibfield  {author} {\bibinfo {author} {\bibfnamefont {S.}~\bibnamefont
  {Tojo}}, \bibinfo {author} {\bibfnamefont {Y.}~\bibnamefont {Taguchi}},
  \bibinfo {author} {\bibfnamefont {Y.}~\bibnamefont {Masuyama}}, \bibinfo
  {author} {\bibfnamefont {T.}~\bibnamefont {Hayashi}}, \bibinfo {author}
  {\bibfnamefont {H.}~\bibnamefont {Saito}}, \ and\ \bibinfo {author}
  {\bibfnamefont {T.}~\bibnamefont {Hirano}},\ }\href {\doibase
  10.1103/PhysRevA.82.033609} {\bibfield  {journal} {\bibinfo  {journal} {Phys.
  Rev. A}\ }\textbf {\bibinfo {volume} {82}},\ \bibinfo {pages} {033609}
  (\bibinfo {year} {2010})}\BibitemShut {NoStop}%
\bibitem [{\citenamefont {Ho}\ and\ \citenamefont {Shenoy}(1996)}]{ho_96}%
  \BibitemOpen
  \bibfield  {author} {\bibinfo {author} {\bibfnamefont {T.-L.}\ \bibnamefont
  {Ho}}\ and\ \bibinfo {author} {\bibfnamefont {V.~B.}\ \bibnamefont
  {Shenoy}},\ }\href {\doibase 10.1103/PhysRevLett.77.3276} {\bibfield
  {journal} {\bibinfo  {journal} {Phys. Rev. Lett.}\ }\textbf {\bibinfo
  {volume} {77}},\ \bibinfo {pages} {3276} (\bibinfo {year}
  {1996})}\BibitemShut {NoStop}%
\bibitem [{\citenamefont {Gautam}\ and\ \citenamefont
  {Angom}(2010{\natexlab{a}})}]{gautam-10}%
  \BibitemOpen
  \bibfield  {author} {\bibinfo {author} {\bibfnamefont {S.}~\bibnamefont
  {Gautam}}\ and\ \bibinfo {author} {\bibfnamefont {D.}~\bibnamefont {Angom}},\
  }\href {http://stacks.iop.org/0953-4075/43/i=9/a=095302} {\bibfield
  {journal} {\bibinfo  {journal} {J. Phys. B}\ }\textbf {\bibinfo {volume}
  {43}},\ \bibinfo {pages} {095302} (\bibinfo {year}
  {2010}{\natexlab{a}})}\BibitemShut {NoStop}%
\bibitem [{\citenamefont {Gautam}\ and\ \citenamefont
  {Angom}(2011)}]{gautam-11}%
  \BibitemOpen
  \bibfield  {author} {\bibinfo {author} {\bibfnamefont {S.}~\bibnamefont
  {Gautam}}\ and\ \bibinfo {author} {\bibfnamefont {D.}~\bibnamefont {Angom}},\
  }\href {http://stacks.iop.org/0953-4075/44/i=2/a=025302} {\bibfield
  {journal} {\bibinfo  {journal} {J. Phys. B}\ }\textbf {\bibinfo {volume}
  {44}},\ \bibinfo {pages} {025302} (\bibinfo {year} {2011})}\BibitemShut
  {NoStop}%
\bibitem [{\citenamefont {Sasaki}\ \emph {et~al.}(2009)\citenamefont {Sasaki},
  \citenamefont {Suzuki}, \citenamefont {Akamatsu},\ and\ \citenamefont
  {Saito}}]{sasaki_09}%
  \BibitemOpen
  \bibfield  {author} {\bibinfo {author} {\bibfnamefont {K.}~\bibnamefont
  {Sasaki}}, \bibinfo {author} {\bibfnamefont {N.}~\bibnamefont {Suzuki}},
  \bibinfo {author} {\bibfnamefont {D.}~\bibnamefont {Akamatsu}}, \ and\
  \bibinfo {author} {\bibfnamefont {H.}~\bibnamefont {Saito}},\ }\href
  {\doibase 10.1103/PhysRevA.80.063611} {\bibfield  {journal} {\bibinfo
  {journal} {Phys. Rev. A}\ }\textbf {\bibinfo {volume} {80}},\ \bibinfo
  {pages} {063611} (\bibinfo {year} {2009})}\BibitemShut {NoStop}%
\bibitem [{\citenamefont {Gautam}\ and\ \citenamefont
  {Angom}(2010{\natexlab{b}})}]{gautam_10}%
  \BibitemOpen
  \bibfield  {author} {\bibinfo {author} {\bibfnamefont {S.}~\bibnamefont
  {Gautam}}\ and\ \bibinfo {author} {\bibfnamefont {D.}~\bibnamefont {Angom}},\
  }\href {\doibase 10.1103/PhysRevA.81.053616} {\bibfield  {journal} {\bibinfo
  {journal} {Phys. Rev. A}\ }\textbf {\bibinfo {volume} {81}},\ \bibinfo
  {pages} {053616} (\bibinfo {year} {2010}{\natexlab{b}})}\BibitemShut
  {NoStop}%
\bibitem [{\citenamefont {Kadokura}\ \emph {et~al.}(2012)\citenamefont
  {Kadokura}, \citenamefont {Aioi}, \citenamefont {Sasaki}, \citenamefont
  {Kishimoto},\ and\ \citenamefont {Saito}}]{kadokura-12}%
  \BibitemOpen
  \bibfield  {author} {\bibinfo {author} {\bibfnamefont {T.}~\bibnamefont
  {Kadokura}}, \bibinfo {author} {\bibfnamefont {T.}~\bibnamefont {Aioi}},
  \bibinfo {author} {\bibfnamefont {K.}~\bibnamefont {Sasaki}}, \bibinfo
  {author} {\bibfnamefont {T.}~\bibnamefont {Kishimoto}}, \ and\ \bibinfo
  {author} {\bibfnamefont {H.}~\bibnamefont {Saito}},\ }\href {\doibase
  10.1103/PhysRevA.85.013602} {\bibfield  {journal} {\bibinfo  {journal} {Phys.
  Rev. A}\ }\textbf {\bibinfo {volume} {85}},\ \bibinfo {pages} {013602}
  (\bibinfo {year} {2012})}\BibitemShut {NoStop}%
\bibitem [{\citenamefont {Takeuchi}\ and\ \citenamefont
  {Kasamatsu}(2013)}]{takeuchi_13}%
  \BibitemOpen
  \bibfield  {author} {\bibinfo {author} {\bibfnamefont {H.}~\bibnamefont
  {Takeuchi}}\ and\ \bibinfo {author} {\bibfnamefont {K.}~\bibnamefont
  {Kasamatsu}},\ }\href {\doibase 10.1103/PhysRevA.88.043612} {\bibfield
  {journal} {\bibinfo  {journal} {Phys. Rev. A}\ }\textbf {\bibinfo {volume}
  {88}},\ \bibinfo {pages} {043612} (\bibinfo {year} {2013})}\BibitemShut
  {NoStop}%
\bibitem [{\citenamefont {Ticknor}(2013)}]{ticknor_13}%
  \BibitemOpen
  \bibfield  {author} {\bibinfo {author} {\bibfnamefont {C.}~\bibnamefont
  {Ticknor}},\ }\href {\doibase 10.1103/PhysRevA.88.013623} {\bibfield
  {journal} {\bibinfo  {journal} {Phys. Rev. A}\ }\textbf {\bibinfo {volume}
  {88}},\ \bibinfo {pages} {013623} (\bibinfo {year} {2013})}\BibitemShut
  {NoStop}%
\bibitem [{\citenamefont {Roy}\ \emph {et~al.}(2014)\citenamefont {Roy},
  \citenamefont {Gautam},\ and\ \citenamefont {Angom}}]{roy_14}%
  \BibitemOpen
  \bibfield  {author} {\bibinfo {author} {\bibfnamefont {A.}~\bibnamefont
  {Roy}}, \bibinfo {author} {\bibfnamefont {S.}~\bibnamefont {Gautam}}, \ and\
  \bibinfo {author} {\bibfnamefont {D.}~\bibnamefont {Angom}},\ }\href
  {\doibase 10.1103/PhysRevA.89.013617} {\bibfield  {journal} {\bibinfo
  {journal} {Phys. Rev. A}\ }\textbf {\bibinfo {volume} {89}},\ \bibinfo
  {pages} {013617} (\bibinfo {year} {2014})}\BibitemShut {NoStop}%
\bibitem [{\citenamefont {Ticknor}(2014)}]{ticknor_14}%
  \BibitemOpen
  \bibfield  {author} {\bibinfo {author} {\bibfnamefont {C.}~\bibnamefont
  {Ticknor}},\ }\href {\doibase 10.1103/PhysRevA.89.053601} {\bibfield
  {journal} {\bibinfo  {journal} {Phys. Rev. A}\ }\textbf {\bibinfo {volume}
  {89}},\ \bibinfo {pages} {053601} (\bibinfo {year} {2014})}\BibitemShut
  {NoStop}%
\bibitem [{\citenamefont {Mason}\ and\ \citenamefont
  {Gardiner}(2014)}]{mason_14}%
  \BibitemOpen
  \bibfield  {author} {\bibinfo {author} {\bibfnamefont {P.}~\bibnamefont
  {Mason}}\ and\ \bibinfo {author} {\bibfnamefont {S.~A.}\ \bibnamefont
  {Gardiner}},\ }\href {\doibase 10.1103/PhysRevA.89.043617} {\bibfield
  {journal} {\bibinfo  {journal} {Phys. Rev. A}\ }\textbf {\bibinfo {volume}
  {89}},\ \bibinfo {pages} {043617} (\bibinfo {year} {2014})}\BibitemShut
  {NoStop}%
\bibitem [{\citenamefont {Hamner}\ \emph {et~al.}(2011)\citenamefont {Hamner},
  \citenamefont {Chang}, \citenamefont {Engels},\ and\ \citenamefont
  {Hoefer}}]{hamner_11}%
  \BibitemOpen
  \bibfield  {author} {\bibinfo {author} {\bibfnamefont {C.}~\bibnamefont
  {Hamner}}, \bibinfo {author} {\bibfnamefont {J.~J.}\ \bibnamefont {Chang}},
  \bibinfo {author} {\bibfnamefont {P.}~\bibnamefont {Engels}}, \ and\ \bibinfo
  {author} {\bibfnamefont {M.~A.}\ \bibnamefont {Hoefer}},\ }\href {\doibase
  10.1103/PhysRevLett.106.065302} {\bibfield  {journal} {\bibinfo  {journal}
  {Phys. Rev. Lett.}\ }\textbf {\bibinfo {volume} {106}},\ \bibinfo {pages}
  {065302} (\bibinfo {year} {2011})}\BibitemShut {NoStop}%
\bibitem [{\citenamefont {\"Ohberg}\ and\ \citenamefont
  {Santos}(2001{\natexlab{a}})}]{ohberg_01}%
  \BibitemOpen
  \bibfield  {author} {\bibinfo {author} {\bibfnamefont {P.}~\bibnamefont
  {\"Ohberg}}\ and\ \bibinfo {author} {\bibfnamefont {L.}~\bibnamefont
  {Santos}},\ }\href {http://stacks.iop.org/0953-4075/34/i=23/a=316} {\bibfield
   {journal} {\bibinfo  {journal} {J. Phys. B}\ }\textbf {\bibinfo {volume}
  {34}},\ \bibinfo {pages} {4721} (\bibinfo {year}
  {2001}{\natexlab{a}})}\BibitemShut {NoStop}%
\bibitem [{\citenamefont {\"Ohberg}\ and\ \citenamefont
  {Santos}(2001{\natexlab{b}})}]{ohberg_01a}%
  \BibitemOpen
  \bibfield  {author} {\bibinfo {author} {\bibfnamefont {P.}~\bibnamefont
  {\"Ohberg}}\ and\ \bibinfo {author} {\bibfnamefont {L.}~\bibnamefont
  {Santos}},\ }\href {\doibase 10.1103/PhysRevLett.86.2918} {\bibfield
  {journal} {\bibinfo  {journal} {Phys. Rev. Lett.}\ }\textbf {\bibinfo
  {volume} {86}},\ \bibinfo {pages} {2918} (\bibinfo {year}
  {2001}{\natexlab{b}})}\BibitemShut {NoStop}%
\bibitem [{\citenamefont {Kevrekidis}\ \emph {et~al.}(2004)\citenamefont
  {Kevrekidis}, \citenamefont {Nistazakis}, \citenamefont {Frantzeskakis},
  \citenamefont {Malomed},\ and\ \citenamefont
  {Carretero-Gonz\'alez}}]{kevrekidis_04}%
  \BibitemOpen
  \bibfield  {author} {\bibinfo {author} {\bibfnamefont {P.~G.}\ \bibnamefont
  {Kevrekidis}}, \bibinfo {author} {\bibfnamefont {H.~E.}\ \bibnamefont
  {Nistazakis}}, \bibinfo {author} {\bibfnamefont {D.~J.}\ \bibnamefont
  {Frantzeskakis}}, \bibinfo {author} {\bibfnamefont {B.~A.}\ \bibnamefont
  {Malomed}}, \ and\ \bibinfo {author} {\bibfnamefont {R.}~\bibnamefont
  {Carretero-Gonz\'alez}},\ }\href {\doibase 10.1140/epjd/e2003-00311-6}
  {\bibfield  {journal} {\bibinfo  {journal} {Eur. Phys. J. D}\ }\textbf
  {\bibinfo {volume} {28}},\ \bibinfo {pages} {181} (\bibinfo {year}
  {2004})}\BibitemShut {NoStop}%
\bibitem [{\citenamefont {Schumayer}\ and\ \citenamefont
  {Apagyi}(2004)}]{schumayer_04}%
  \BibitemOpen
  \bibfield  {author} {\bibinfo {author} {\bibfnamefont {D.}~\bibnamefont
  {Schumayer}}\ and\ \bibinfo {author} {\bibfnamefont {B.}~\bibnamefont
  {Apagyi}},\ }\href {\doibase 10.1103/PhysRevA.69.043620} {\bibfield
  {journal} {\bibinfo  {journal} {Phys. Rev. A}\ }\textbf {\bibinfo {volume}
  {69}},\ \bibinfo {pages} {043620} (\bibinfo {year} {2004})}\BibitemShut
  {NoStop}%
\bibitem [{\citenamefont {Dodd}\ \emph {et~al.}(1998)\citenamefont {Dodd},
  \citenamefont {Edwards}, \citenamefont {Clark},\ and\ \citenamefont
  {Burnett}}]{dodd_98}%
  \BibitemOpen
  \bibfield  {author} {\bibinfo {author} {\bibfnamefont {R.~J.}\ \bibnamefont
  {Dodd}}, \bibinfo {author} {\bibfnamefont {M.}~\bibnamefont {Edwards}},
  \bibinfo {author} {\bibfnamefont {C.~W.}\ \bibnamefont {Clark}}, \ and\
  \bibinfo {author} {\bibfnamefont {K.}~\bibnamefont {Burnett}},\ }\href
  {\doibase 10.1103/PhysRevA.57.R32} {\bibfield  {journal} {\bibinfo  {journal}
  {Phys. Rev. A}\ }\textbf {\bibinfo {volume} {57}},\ \bibinfo {pages} {R32}
  (\bibinfo {year} {1998})}\BibitemShut {NoStop}%
\bibitem [{\citenamefont {Griffin}(1996)}]{griffin_96}%
  \BibitemOpen
  \bibfield  {author} {\bibinfo {author} {\bibfnamefont {A.}~\bibnamefont
  {Griffin}},\ }\href {\doibase 10.1103/PhysRevB.53.9341} {\bibfield  {journal}
  {\bibinfo  {journal} {Phys. Rev. B}\ }\textbf {\bibinfo {volume} {53}},\
  \bibinfo {pages} {9341} (\bibinfo {year} {1996})}\BibitemShut {NoStop}%
\bibitem [{\citenamefont {Muruganandam}\ and\ \citenamefont
  {Adhikari}(2009)}]{muruganandam_09}%
  \BibitemOpen
  \bibfield  {author} {\bibinfo {author} {\bibfnamefont {P.}~\bibnamefont
  {Muruganandam}}\ and\ \bibinfo {author} {\bibfnamefont {S.~K.}\ \bibnamefont
  {Adhikari}},\ }\href {doi:10.1016/j.cpc.2009.04.015} {\bibfield  {journal}
  {\bibinfo  {journal} {Comp. Phys. Comm.}\ }\textbf {\bibinfo {volume}
  {180}},\ \bibinfo {pages} {1888} (\bibinfo {year} {2009})}\BibitemShut
  {NoStop}%
\bibitem [{\citenamefont {Anderson}\ \emph {et~al.}(1999)\citenamefont
  {Anderson}, \citenamefont {Bai}, \citenamefont {Bischof}, \citenamefont
  {Blackford}, \citenamefont {Demmel}, \citenamefont {Dongarra}, \citenamefont
  {Croz}, \citenamefont {Greenbaum}, \citenamefont {Hammarling}, \citenamefont
  {McKenney},\ and\ \citenamefont {Sorensen}}]{anderson_99}%
  \BibitemOpen
  \bibfield  {author} {\bibinfo {author} {\bibfnamefont {E.}~\bibnamefont
  {Anderson}}, \bibinfo {author} {\bibfnamefont {Z.}~\bibnamefont {Bai}},
  \bibinfo {author} {\bibfnamefont {C.}~\bibnamefont {Bischof}}, \bibinfo
  {author} {\bibfnamefont {S.}~\bibnamefont {Blackford}}, \bibinfo {author}
  {\bibfnamefont {J.}~\bibnamefont {Demmel}}, \bibinfo {author} {\bibfnamefont
  {J.}~\bibnamefont {Dongarra}}, \bibinfo {author} {\bibfnamefont {J.~D.}\
  \bibnamefont {Croz}}, \bibinfo {author} {\bibfnamefont {A.}~\bibnamefont
  {Greenbaum}}, \bibinfo {author} {\bibfnamefont {S.}~\bibnamefont
  {Hammarling}}, \bibinfo {author} {\bibfnamefont {A.}~\bibnamefont
  {McKenney}}, \ and\ \bibinfo {author} {\bibfnamefont {D.}~\bibnamefont
  {Sorensen}},\ }\href@noop {} {\emph {\bibinfo {title} {{LAPACK} Users'
  Guide}}},\ \bibinfo {edition} {3rd}\ ed.\ (\bibinfo  {publisher} {Society for
  Industrial and Applied Mathematics},\ \bibinfo {address} {Philadelphia, PA},\
  \bibinfo {year} {1999})\BibitemShut {NoStop}%
\bibitem [{\citenamefont {G\"orlitz}\ \emph {et~al.}(2001)\citenamefont
  {G\"orlitz}, \citenamefont {Vogels}, \citenamefont {Leanhardt}, \citenamefont
  {Raman}, \citenamefont {Gustavson}, \citenamefont {Abo-Shaeer}, \citenamefont
  {Chikkatur}, \citenamefont {Gupta}, \citenamefont {Inouye}, \citenamefont
  {Rosenband},\ and\ \citenamefont {Ketterle}}]{gorlitz_01}%
  \BibitemOpen
  \bibfield  {author} {\bibinfo {author} {\bibfnamefont {A.}~\bibnamefont
  {G\"orlitz}}, \bibinfo {author} {\bibfnamefont {J.~M.}\ \bibnamefont
  {Vogels}}, \bibinfo {author} {\bibfnamefont {A.~E.}\ \bibnamefont
  {Leanhardt}}, \bibinfo {author} {\bibfnamefont {C.}~\bibnamefont {Raman}},
  \bibinfo {author} {\bibfnamefont {T.~L.}\ \bibnamefont {Gustavson}}, \bibinfo
  {author} {\bibfnamefont {J.~R.}\ \bibnamefont {Abo-Shaeer}}, \bibinfo
  {author} {\bibfnamefont {A.~P.}\ \bibnamefont {Chikkatur}}, \bibinfo {author}
  {\bibfnamefont {S.}~\bibnamefont {Gupta}}, \bibinfo {author} {\bibfnamefont
  {S.}~\bibnamefont {Inouye}}, \bibinfo {author} {\bibfnamefont
  {T.}~\bibnamefont {Rosenband}}, \ and\ \bibinfo {author} {\bibfnamefont
  {W.}~\bibnamefont {Ketterle}},\ }\href {\doibase
  10.1103/PhysRevLett.87.130402} {\bibfield  {journal} {\bibinfo  {journal}
  {Phys. Rev. Lett.}\ }\textbf {\bibinfo {volume} {87}},\ \bibinfo {pages}
  {130402} (\bibinfo {year} {2001})}\BibitemShut {NoStop}%
\bibitem [{\citenamefont {Pattinson}\ \emph {et~al.}(2013)\citenamefont
  {Pattinson}, \citenamefont {Billam}, \citenamefont {Gardiner}, \citenamefont
  {McCarron}, \citenamefont {Cho}, \citenamefont {Cornish}, \citenamefont
  {Parker},\ and\ \citenamefont {Proukakis}}]{pattinson_13}%
  \BibitemOpen
  \bibfield  {author} {\bibinfo {author} {\bibfnamefont {R.~W.}\ \bibnamefont
  {Pattinson}}, \bibinfo {author} {\bibfnamefont {T.~P.}\ \bibnamefont
  {Billam}}, \bibinfo {author} {\bibfnamefont {S.~A.}\ \bibnamefont
  {Gardiner}}, \bibinfo {author} {\bibfnamefont {D.~J.}\ \bibnamefont
  {McCarron}}, \bibinfo {author} {\bibfnamefont {H.~W.}\ \bibnamefont {Cho}},
  \bibinfo {author} {\bibfnamefont {S.~L.}\ \bibnamefont {Cornish}}, \bibinfo
  {author} {\bibfnamefont {N.~G.}\ \bibnamefont {Parker}}, \ and\ \bibinfo
  {author} {\bibfnamefont {N.~P.}\ \bibnamefont {Proukakis}},\ }\href {\doibase
  10.1103/PhysRevA.87.013625} {\bibfield  {journal} {\bibinfo  {journal} {Phys.
  Rev. A}\ }\textbf {\bibinfo {volume} {87}},\ \bibinfo {pages} {013625}
  (\bibinfo {year} {2013})}\BibitemShut {NoStop}%
\bibitem [{\citenamefont {Lercher}\ \emph {et~al.}(2011)\citenamefont
  {Lercher}, \citenamefont {Takekoshi}, \citenamefont {Debatin}, \citenamefont
  {Schuster}, \citenamefont {Rameshan}, \citenamefont {Ferlaino}, \citenamefont
  {Grimm},\ and\ \citenamefont {Nägerl}}]{lercher_11}%
  \BibitemOpen
  \bibfield  {author} {\bibinfo {author} {\bibfnamefont {A.}~\bibnamefont
  {Lercher}}, \bibinfo {author} {\bibfnamefont {T.}~\bibnamefont {Takekoshi}},
  \bibinfo {author} {\bibfnamefont {M.}~\bibnamefont {Debatin}}, \bibinfo
  {author} {\bibfnamefont {B.}~\bibnamefont {Schuster}}, \bibinfo {author}
  {\bibfnamefont {R.}~\bibnamefont {Rameshan}}, \bibinfo {author}
  {\bibfnamefont {F.}~\bibnamefont {Ferlaino}}, \bibinfo {author}
  {\bibfnamefont {R.}~\bibnamefont {Grimm}}, \ and\ \bibinfo {author}
  {\bibfnamefont {H.-C.}\ \bibnamefont {Nägerl}},\ }\href {\doibase
  10.1140/epjd/e2011-20015-6} {\bibfield  {journal} {\bibinfo  {journal} {Euro.
  Phys. Jour. D}\ }\textbf {\bibinfo {volume} {65}},\ \bibinfo {pages} {3}
  (\bibinfo {year} {2011})}\BibitemShut {NoStop}%
\bibitem [{\citenamefont {Pilch}\ \emph {et~al.}(2009)\citenamefont {Pilch},
  \citenamefont {Lange}, \citenamefont {Prantner}, \citenamefont {Kerner},
  \citenamefont {Ferlaino}, \citenamefont {N\"agerl},\ and\ \citenamefont
  {Grimm}}]{pilch_09}%
  \BibitemOpen
  \bibfield  {author} {\bibinfo {author} {\bibfnamefont {K.}~\bibnamefont
  {Pilch}}, \bibinfo {author} {\bibfnamefont {A.~D.}\ \bibnamefont {Lange}},
  \bibinfo {author} {\bibfnamefont {A.}~\bibnamefont {Prantner}}, \bibinfo
  {author} {\bibfnamefont {G.}~\bibnamefont {Kerner}}, \bibinfo {author}
  {\bibfnamefont {F.}~\bibnamefont {Ferlaino}}, \bibinfo {author}
  {\bibfnamefont {H.-C.}\ \bibnamefont {N\"agerl}}, \ and\ \bibinfo {author}
  {\bibfnamefont {R.}~\bibnamefont {Grimm}},\ }\href {\doibase
  10.1103/PhysRevA.79.042718} {\bibfield  {journal} {\bibinfo  {journal} {Phys.
  Rev. A}\ }\textbf {\bibinfo {volume} {79}},\ \bibinfo {pages} {042718}
  (\bibinfo {year} {2009})}\BibitemShut {NoStop}%
\bibitem [{\citenamefont {McCarron}\ \emph {et~al.}(2011)\citenamefont
  {McCarron}, \citenamefont {Cho}, \citenamefont {Jenkin}, \citenamefont
  {K\"oppinger},\ and\ \citenamefont {Cornish}}]{mccarron_11}%
  \BibitemOpen
  \bibfield  {author} {\bibinfo {author} {\bibfnamefont {D.~J.}\ \bibnamefont
  {McCarron}}, \bibinfo {author} {\bibfnamefont {H.~W.}\ \bibnamefont {Cho}},
  \bibinfo {author} {\bibfnamefont {D.~L.}\ \bibnamefont {Jenkin}}, \bibinfo
  {author} {\bibfnamefont {M.~P.}\ \bibnamefont {K\"oppinger}}, \ and\ \bibinfo
  {author} {\bibfnamefont {S.~L.}\ \bibnamefont {Cornish}},\ }\href {\doibase
  10.1103/PhysRevA.84.011603} {\bibfield  {journal} {\bibinfo  {journal} {Phys.
  Rev. A}\ }\textbf {\bibinfo {volume} {84}},\ \bibinfo {pages} {011603}
  (\bibinfo {year} {2011})}\BibitemShut {NoStop}%
\end{thebibliography}%
\bibliographystyle{apsrev4-1}

\end{document}